\pdfoutput=1
\documentclass[11pt]{article}
\usepackage{amssymb}
\usepackage{graphicx}
\usepackage{array}
\parskip \baselineskip
\newcommand{\ds}{\displaystyle}
\newcommand{\R}{\mathbb{R}}

\newcommand{\s}{\subseteq}

\newcommand{\ol}{\overline}

\newcommand{\ra}{\rightarrow}
\newcommand{\Ra}{\Rightarrow}
\newcommand{\es}{\emptyset}
\newcommand{\LRa}{\Leftrightarrow}
\newcommand{\name}[2]{{#1}{\scriptsize{#2}}}

\begin{document}
\begin{center}
{\bf Compression with wildcards: All, or all maximum, anticliques of  graph}

\name{M}{arcel} \name{W}{ild}
\end{center}
\begin{quote}
A{\scriptsize BSTRACT:} {\footnotesize By definition an anticlique is an independent set of vertices of a graph $G$. By duality all results obtained for anticliques carry over to cliques.
(It is for technical reasons that we stick with anticliques throughout.) We display the set $Acl(G)$ of all anticliques of $G$ in a compressed format that uses wildcards.  Likewise (albeit less compressed) for the subfamily $MACL(G)\s Acl(G)$ of all maximum-cardinality members. The second task works particularly well for bipartite graphs (in fact for the broader class of K\"onig-Egarv\'ary graphs). In  this scenario Boolean functions (of type 2-CNF) will be important. Posets and lattices also feature prominently. }
\end{quote}


\section{Introduction}

Throughout the article, and without further mention, all structures will be finite. Always $G=(V,E)$ is a simple graph with vertex-set $V$ and edge-set\footnote{Formally $E$ is a set of 2-element sets. Thus if $\{a,b\}\in E$ then $\{a,b\}$ is an edge that "connects" vertex $a$ to vertex $b$.} $E$. Vertices $a,b$ are {\it adjacent } if $\{a,b\}\in E$. Recall that a {\it clique} is a set $X\s V$ such that any two vertices in $X$ are adjacent.
In contrast,  an {\it anticlique} $X\s V$ (also known as {\it independent} set or {\it stable} set) is such that no two vertices in $X$ are adjacent. Our main quest is to {\it compress} the set $Acl(G)$ of all anticliques of $G$, using novel types of wildcards.

Recall that the {\it dual} graph $G^d$ of $G=(V,E)$ has the same vertex-set  and its edges are the "non-edges" of $G$.
 Formally $G^d=(V,E^d)$, where $E^d:=\big\{\{a,b\}:\ a,b\in V\ \wedge\ a\neq b\ \wedge\ \{a,b\}\not\in E\big\}$.
It follows that the cliques  of $G$ coincide with the anticliques  of
 $G^d$, and so all our results about anticliques carry over to cliques. In the sequel we stick however to anticliques since 
  the technicalities of our main algorithm require anticliques rather than cliques. Many concepts in discrete mathematics can be viewed as the anticliques (or cliques) of some "auxiliary" graph. For instance the antichains of a poset $P$ are the anticliques of the comparability graph of $P$.  The reader may ask: Why targeting the {\it whole} of $Acl(G)$ in the first place? Isn't it  sufficient to know the {\it maximal} anticliques? 
  
  Not always. For instance,  the {\it Merriefield-Simmons index} $ms(G):=|Acl(G)|$ is a crucial parameter in chemistry [WG]. 
In another vein, take a
weighting function $wt:V\to \R^+$ and (as often) for $X\s V$ define $wt(X)$ as the sum of all $wt(x)\ (x\in X)$. It is obvious that an anticlique of maximum value $wt(Y)$ necessarily is inclusion-maximal. However this needs not be the case for more subtle weighting functions such as $wt'(X):=wt(X)-|X|^2$. More examples later.

Granting that the whole of $Acl(G)$ may be important, one may still wonder:  Why not calculating $Acl(G)$ as a union of powersets, i.e. as $Acl(G)=P(F_1)\cup P(F_2)\cup\cdots \cup P(F_s)$ where $F_1,...,F_s$ are all maximal anticliques? Because first the $F_i$'s need to be calculated. Second (and worse), there is a lot of overlap among the different powersets, which makes the representation next to useless and which is hard to correct, see [W3].
  
  Of course, for some purposes the maximal anticliques {\it do} suffice. Since many algorithms exist to generate them all, we won't walk that beaten path - but a related one. Specifically, putting 
  $$\alpha(G):=max\{|X|:\ X\in Acl(G)\}$$
  we  strive to capture the structure of
  the family
$$(1)\quad MACL(G):=\{X\in Acl(G):\ |X|=\alpha(G)\}$$
of all {\it maximum} (cardinality) anticliques.
While a few results (Lemma 2, Lemma 3) can be obtained for arbitrary graphs $G$, we mainly focus on $MACL(G)$ for so called K\"onig-Egarv\'ary graphs (which comprise all bipartite graphs).  Furthermore, we propose two methods to reduce an {\it arbitrary} graph $G$ to bipartite constituents $B_i\ (i\le R)$. Here "reduce" means that $Acl(G)$ is the union of the families $Acl(B_i)\ (i\le R)$, and $MACL(G)$ is the union of the families $MACL(B_i)\ (i\le R)$. Each family $Acl(B_i)$ or $MACL(B_i)$ is readily found. The first method is the $(a,c)$-algorithm, the second is the bip-val algorithm (the acronym meaning bipartite-valued).

\vspace{0.5cm}
{\bf 1.1} Here comes the Section break up.
Section 2 presents the {\it } $(a,c)${\it -algorithm} which for arbitrary $G$
generates all its anticliques. Yet not one-by-one. Instead it partitions $Acl(G)$ into certain 012ac-rows (or simply rows)  $r_i\s Acl(G)$,  each one of which potentially encoding a large number $N(r_i)$ of anticliques. Hence the number $R$ of  rows $r_i$  can be much smaller than $N:=|Acl(G)|$. All of this works (Thm.1) in output polynomial time $O(Rw^2)=O(Nw^2)$. Section 3 introduces Boolean 2-CNFs (i.e. Boolean functions in Conjunctive Normal Form all of whose clauses have length two) and their relation to $Acl(G)$. Section 4 widens the horizon by briefly zooming out from graphs to hypergraphs. For instance what is a hitting-set on the hypergraph level becomes  vertex-cover on the graph level. Furthermore Section 4 introduces the complementary concepts of  anticlique and  vertex-cover, and recalls that a   K\"oenig-Egarv\'ary  graph is such that the size of a maximum matching equals the size of a minimum vertex-cover.

In Section 5 our compression of $Acl(G)$  yields   $MACL(G)$ in straightforward ways, albeit not in polynomial total time.
Section 6 offers a new approach, based on $Acl(G)=r_1\uplus\cdots\uplus r_R$, to the independence polynomial $I(G,x)=\sum_{k=0}^\alpha s_k x^k$, where $\alpha:=\alpha(G)$ and where by definition $s_k$ is the number of $k$-element anticliques of the graph $G$.
Section 7 tackles the {\it generation} (as opposed to counting) of $k$-anticliques.

If $G=(V,E)$ and $S\s V$ then $G[S]$ is the subgraph induced by $S$.
Section 8 introduces the $(V_1,V_2)$-approach, which aims to enhance the (a,c)-algorithm. The idea is to split $V$ as $V=V_1\uplus V_2$ in such a way that $Acl(G[V_1])$ is small enough to be enumerated one-by-one, but $Acl(G[V_2])$ can be gigantic as long as it can be compressed with few 012ac-rows. The better the individual qualities of $V_1,V_2$, the better one can compress $Acl(G)$ itself. The quest to find quality subsets $V_1,V_2$ leads to the well known concept of an odd-cycle-cover $V_1$, which by definition forces $Acl(G[V\setminus V_1])$ to be bipartite. Therefore, as desired, $Acl(G[V_2])\ (V_2:=V\setminus V_1)$ can be compressed particularly well. If one wishes to trim "bipartite" to "forest", this can be achieved by trimming "odd-cycle-cover" to "feedback-vertex-set". 

  As to Section 9, it is long known that the minimum vertex-covers (=mvcs) of a  K\"oenig-Egarv\'ary {\it with} a perfect matching and the models of a pure 2-CNF (a type of Boolean function) are essentially the "same thing". We reprove this in a fresh way in Theorem 5. This leads, in Section 10, to new insights about the join-irreducibles of the Dilworth lattice $Dil(P,\le)$, i.e. the lattice of all maximum antichains of
  $(P,\le)$. Theorem 5 also is the key, in Section 11, to unravel the structure of $MACL(G)$ when $G$ is K\"oenig-Egarv\'ary {\it without} perfect matching.

 The $(V_1,V_2)$-approach in Section 8 is one way to step up the (a,c)-algorithm; albeit in all of that 012ac-rows remain the basic ingredients.
 In contrast, Section 12 generalizes the (a,c)-algorithm and its 012ac-rows to a certain bip-val algorithm and its accompanying  {\it bipartite-valued} rows. We go on to investigate the pros and cons of the two methods.

\section{The  (a,c)-algorithm}

{\bf 2.1} Consider the graph $G_1$ with vertex-set $V = [6]:=\{1,2,...,6\}$ in Fig.1 (and ignore the coloring of vertices). The anticliques $X \in Acl(G_1)$ with $1 \not\in X$ are exactly\footnote{ More precisely the $X$'s are the anticliques in the {\it graph}  $G[V\backslash \{1\}]$ induced by the set $V\setminus \{1\}$. Recall, if $H=(V,E)$ and $S\s V$ then $H[S]$ is the graph with vertex-set $S$ whose edges are exactly the edges $\{u,v\}\in E$ satisfying $u,v\in S$.}
the anticliques in $V\setminus\{1\}$. On the other hand, an anticlique $X$ of $G_1$ with $1 \in X$ does not contain the neighbours $2, 4, 5$ of $1$. 

\begin{center}
\includegraphics[scale=1.1]{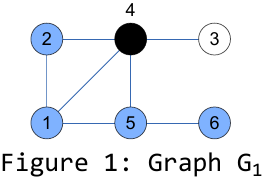}
\end{center}

It follows that $Acl(G_1)$ is contained in the  disjoint union of the set families $r_1$ and $r_2$ in Table 1. Here we identify subsets of $V$ with bitstrings in the usual way and let the {\it don't-care} symbol 2 freely assume the values 0 and 1.

\begin{tabular}{c|c|c|c|c|c|c|l}
& 1 & 2 & 3 & 4 & 5 & 6\\ \hline
 &  &   &   &   &   & \\ \hline
$r_1=$ & ${\bf 0}$ & 2 & 2 & 2 & 2 & 2 & pending 2 \\ \hline
$r_2=$ & ${\bf 1}$ & 0 & 2 & 0 & 0& 2 & final \\ \hline
&  &   &   &   &   & \\ \hline

$r_3=$ & 0 & ${\bf 0}$ & 2 & 2 & 2 & 2 & pending 3\\ \hline
$r_4=$ & 0 & ${\bf 1}$ & 2 & 0 & 2 & 2 & pending 5 \\ \hline
&  &   &   &   &   & \\ \hline

$r_5 =$ & 0 & 0 & ${\bf 0}$ & 2 & 2 & 2 & pending 4\\ \hline
$r_6=$ & 0 & 0 & ${\bf 1}$ &  0 & 2 & 2 & pending 5\\ \hline
$r_4=$ & 0 & 1 & 2 & 0 & 2 & 2 & pending 5 \\ \hline
&  &   &  &   &  & \\ \hline

$r_7 =$ & 0 & 0 & 0 & 0 & 2 & 2 & pending 5\\ \hline
$r_6=$ & 0 & 0 & 1 &  0 & 2 & 2 & pending 5\\ \hline
$r_4=$ &0 & 1 & 2 & 0 & 2 & 2 & pending 5\\ \hline 
\end{tabular} 
 \hspace*{2cm} 
\begin{tabular}{c|c|c|c|c|c|c|} 
 &  1& 2 & 3 & 4 & 5 & 6\\ \hline
  &  &   &   &   &   & \\ \hline
  $r_{14}=$ & 0 & 1 & 2 & 0 & 1 & 0 \\ \hline
  $r_{13}=$ & 0& 1 & 2 & 0 & 0 & 2\\ \hline
  $r_{12}=$ & 0 & 0 & 1 & 0 & 1 & 0 \\ \hline
  $r_{11}=$ & 0 & 0 & 1 & 0 & 0 & 2\\ \hline
  $r_{10} =$ & 0 & 0& 0 & 0 & 1 & 0\\ \hline
  $r_9=$ & 0 & 0 & 0 & 0 & 0 & 2 \\ \hline
  $r_8 =$ & 0 & 0 & 0 & 1& 0 & 2\\ \hline
  $r_2 =$ & 1 & 0 & 2 & 0 & 0 & 2 \\ \hline \end{tabular}
  
  {\sl Table 1: The working stack \hspace*{4.5cm} Table 2: The final stack}
	
	We refer to set families like $r_1$ or $r_2$ as {\it 012-rows}. Similarly every anticlique (=acl) of $G_1$ that contains $2$ must\footnote{Notice that cliques cannot be dealt with in analogous fashion: A clique $X$ of some graph $G$ with $a\in X$ {\it need not} contain all neighbors of $a$. As shown in Section 1, the tool that takes cliques on board is the dual graph $G^d$.} not contain $1,\, 4$; 
	in brief $2 \ra \ol{1,4}$. The corresponding treatment of vertex $2$ is {\it pending} in $r_1$. As to row $r_2$, the four sets  it contains happen to be acls, and so $r_2$ becomes the first member of the {\it final stack} in Table 2. Imposing the pending {\it quasi-implication}\footnote{The precise relation to Boolean functions is postponed to Section 3.}
$2 \ra \ol{1,4}$ upon $r_1$ replaces it by the disjoint union of $r_3$ and $r_4$. Notice that $r_4$ satisfies both $3 \ra \ol{4}$  and $4 \ra \ol{1,2,3,5}$ (both because $4\not\in X$ for all $X\in r_4$). Consequently vertex $5$ is pending in $r_4$. 

Imposing $3 \ra \ol{4}$ on the top row $r_3$ yields $r_5$ and $r_6$. Imposing $4 \ra \ol{1,2,3,5}$ on the new top row $r_5$ yields $r_7$ and the final row $r_8$ (Table 2). Imposing
 $5 \ra \ol{1,4,6}$ upon the rows $r_7,\ r_6,\ r_4$ yields the final rows $r_9$ to $r_{14}$ and empties the working stack. The eight final rows in Table 2 yield a total of $ms(G_1)=|r_{14}|+|r_{13}|+\cdots+|r_{2}|=2+4+\cdots+4=18$ anticliques.

{\bf 2.2} One calls $Z\s V$ is a {\it vertex-cover} of $G=(V,E)$ if $Z\cap\{a,b\}\neq\es$ for all $\{a,b\}\in E$. 
in plain words, each edge of $G$ must be "touched" by at least one vertex in $Z$.
Vertex-covers $Z$  smaller than the trivial vertex-cover $V$ will be useful, because it suffices that the vertices in $Z$ are fed to the (a,c)-algorithm. This is carried out for $G_1$ and $Z:=\{1,4,6\}$ in
 Table 3 (we omitted the final stack). One checks that indeed 
$|\rho_{1}|+|\rho_{2}|+|\rho_{3}|+|\rho_{4}|=18$.

\begin{tabular}{c|c|c|c|c|c|c|l}
& 1 & {\bf 2} & 3 & {\bf 4} & 5 & {\bf 6} \\ \hline
 &  &   &   &   &   & \\ \hline

           & ${\bf 0}$ & 2 & 2 & 2 & 2 & 2 & p.4 \\ \hline
$\rho_1=$ & ${\bf 1}$ & 0 & 2 & 0 & 0& 2 & final\\ \hline
&  &   &   &   &   & \\ \hline

          & 0 & 2 & 2 & ${\bf 0}$ & 2 & 2 & p.6\\ \hline
$\rho_2=$ & 0 & 0 & 0 & ${\bf 1}$ & 0 & 2 & final \\ \hline
&  &   &   &   &   & \\ \hline

$\rho_3=$ & 0 & 2 & 2 &  0 & 2 & ${\bf 0}$ & final\\ \hline
$\rho_4=$ &0 & 2 & 2 & 0 & 0 & ${\bf 1}$ & final\\ \hline 
\end{tabular}

{\sl Table 3: Shrinking the working stack by virtue of the vertex-cover $\{1,4,6\}$}
 
 {\bf 2.3} As observed in 2.2, $Acl(G_1)$ is the family of all $X \subseteq [6]$ that  simultaneously satisfy the three quasi-implications induced by the vertex-cover $\{4, 1, 6\}$. (That particular ordering will further speed up matters.) Small vertex-covers  perform even better when accompanied by the use of wildcards fancier than the don't-care symbol 2. 
 
 Formally an {\it (a,c)-wildcard} is a certain collection of bitstrings:
 $$(2)\quad (a,c,c,...,c):=({\bf 0},2,2,...,2)\uplus ({\bf 1},0,0,...,0)$$
 It follows that row $\sigma_1$ in Table 4 encodes the set family ${\cal F}$ all $X \subseteq [6]$ that satisfy\footnote{The meaning of "satisfy" will be rendered more precisely in Section 3. } the quasi-implication $4 \ra \ol{1,2,3,5}$. Thus $X \in {\cal F}$ iff either $4 \not\in X$ or $1, 2, 3, 5 \not\in X$ (or both). 
 
 \begin{tabular}{c|c|c|c|c|c|c|l} \hline
 & 1 & 2& 3 & 4 & 5 & 6\\ \hline
 &   &  &   &   &   & \\ \hline
 $\sigma_1=$ & $c$ & $c$ & $c$ & $a$ & $c$ & 2 & p.1\\ \hline
 &   &  &   &   &   & \\ \hline
 $\sigma_2=$ & ${\bf 0}$ & $c$ & $c$ & $a$ & $c$ & 2 &p.6\\ \hline
 $\sigma_3=$ & ${\bf 1}$ & 0 & 2 & 0 & 0 & 2 & final\\ \hline 
&   &  &   &   &   & \\ \hline
$\sigma_4=$ & 0 & $c$ & $c$ & $a$ & $c$ & ${\bf 0}$ & final\\ \hline
$\sigma_5=$ & 0 & $c$ & $c$ & $a$ & 0 & ${\bf 1}$ & final \\ \hline
\end{tabular} 

 {\sl Table 4: Using wildcards further compresses the working stack} 
 
 In order to sieve those $X \in \sigma_1$ that satisfy $1 \ra \ol{2,4,5}$ we again distinguish two kinds of $X$; the ones with $1 \not\in X$ and the ones with $1 \in X$. Obviously the first kind constitutes $\sigma_2$. More subtle, the second kind constitutes $\sigma_3$. 
Namely, switching the first $c$ in $\sigma_1$ to $1$  forces $a=0$, which in turn 'releases' the three remaining $c$'s. Thus one gets the preliminary row
$(1,2,2,0,2,2)$. Inforcing $1 \ra \ol{2,4,5}$ on it yields $\sigma_3$. It happens that $\sigma_3$  is final.
The top row $\sigma_2$ of the working stack gives way, upon imposing $6 \ra \ol{5}$ on it, to the final rows $\sigma_4$ and $\sigma_5$ in Table 4. 
 One verifies\footnote{Generally (2) implies that $|(a,c,...,c)|=2^t+1$ where $t$ is the number of $c$'s.} that $|\sigma_3| + |\sigma_4| + |\sigma_5|= 4 + 9 + 5 = 18$. This matches the previously obtained number, but uses  fewer rows still. 
 
{\bf 2.4} To take a more telling example, for a random graph with 35 vertices and 60 edges the method of Subsection 2.1, call it {\it $012$-algorithm}, needed 6.6 seconds to pack the
$4'046'882$ anticliques into 53669 final 012-rows. Computing a minimum vertex-cover of cardinality 18 took less than $0.01$ sec (using Mathematica) and the so adapted 012-algorithm produced only $25152$ final 012-rows and needed 1.1 seconds.  Additionally using the (a,c)-wildcard resulted in a mere $1060$ final 012ac-rows and took 0.2 seconds. (We henceforth call this method the {(a,c)-algorithm}.) The drops $53669>25125>1060$ with corresponding times $6.6>1.1>0.2$ are drastic. The Mathematica command {\tt BooleanConvert} renders the modelset of any Boolean function as disjoint\footnote{Commonly used synonyms for "disjoint union of 012-row" are "Exclusive sum of products (ESOP)" and "Orthogonal DNF".} union of 012-rows, in particular (see 2.6) it is fit to compress $Acl(G)$. However, the (a,c)-algorithm is superior.

{\bf 2.5} Full details on  the (a,c)-algorithm, are provided in the proof of Theorem 1 below. There, for the sake of notation, we process the whole vertex-set $[w]$, instead\footnote{Using $Z$ instead of $[w]$ would not improve the $O(w^2)$ bound in Theorem 1.} of a vertex-cover $Z\subseteq [w]$.

{\bf Theorem 1: }{\it Let $G$ be a graph with vertex-set $[w]$. 
Then the (a,c)-algorithm generates the $N:=ms(G)$ anticliques in time $O(Nw^2)$, and in a compressed format.} 

{\it Proof.} We impose our quasi-implications in the order $1 \ra \ol{NH(1)},  \cdots, w \ra \ol{NH(w)}$, where $NH(i)$ is the neighbourhood of vertex $i$.  Consider a ``generic'' top row $r$ of the working stack which has $t \ra \ol{C}$ as {\it pending} anti-implication, where $C = NH(t)$. (This means that all anti-implications $i \ra \ol{NH(i)}$ with $1\le i<t$ have been imposed on $r$ already.) In order to impose $t \ra \ol{C}$ upon $r$ we assume by induction that the symbol $\sigma$ at position $t$ in row $r$ is either $0,2$ or $c$ (but neither $a$ nor $1$). These will be our three main cases. If we think of the vertices as being processed from ``left to right'', the situation is as follows:
$$r\quad =\quad( \cdots \underbrace{\cdots \cdots,}_{C} \underbrace{\sigma ,}_{t} \underbrace{\cdots \cdots,}_{C} \cdots \cdots )$$
Our inductive assumption about $a$ and $1$ is admissible provided that we never write $a$ or $1$ to the right of $\sigma$ in the upcoming process of changing row $r$.

{\sl Case 1}: $\sigma =0$ or the whole of $C$ consists of $0$'s. Then $t \ra \ol{C}$ is satisfied by all $X \in r$, so $r$ survives unaltered.

In the following cases there cannot be a symbol 1 in $C$ (also not to the  left of $\sigma$) since then $\sigma$, being a neighbour of the 1-labelled vertex, would have been set to $0$. (Thus the only $1$'s in $C$ occur in Case 1 when $\sigma =0$, and then they are to the left of $\sigma$.)

{\sl Case 2}: $\sigma =2$.  Then one of the subcases 2.1 or 2.2 takes place.

Subcase 2.1: $C$ consists only of $0$'s and of $2$'s (with at least one $2$  by Case 1). Then put $\sigma =a$ and put $c$ on all the $2$'s within $C$. 

Subcase 2.2: There is at least one $a$ or $c$ on a $C$-position. Consider the row

 $$r\quad =\quad (\cdots, 1, c_3, a_4, a_3, \underbrace{a_1, a_2, c_2,}_{C} 
\underbrace{\bf 2,}_{t} \underbrace{c_3, c_4, 0, 2,}_{C} c_2, c_1, 0, \cdots ),$$

which is typical in that it illustrates all four possibilities: from the wildcard $a_1c_1$ only the premise $a_1$ is in $C$; from\footnote{We sometimes use shortcuts like $a_2c_2c_2$ for $(a_2,c_2,c_2)$.} $a_2c_2c_2$ the premise and a nonempty part (possibly all) of the "conclusion" $c_2c_2$ is in $C$; from $a_3c_3 c_3$ only a proper part of the conclusion is in $C$; and from $a_4c_4$ only the complete conclusion is in $C$.
A moment's thought shows that the sets $X \in r$ that satisfy $t \ra \ol{C}$ are precisely the ones in the (disjoint) union of $r'$ and $r''$, where
$$\begin{array}{lllllllllllllllllll} r' & = & (\cdots 1, & c_3, & a_4, & a_3, & a_1, & a_2, & c_2, & {\bf 0}, & c_3, & c_4, & 0,  & 2, & c_2, & c_1, & 0, & \cdots ),\\
\\
r'' & = & (\cdots 1, & c_3, & 2, & a_3, & 0, & 0, & 0, & {\bf 1}, & 0, & 0, & 0, & 0, & 2, & 2, & 0, & \cdots). \end{array}$$

{\sl Case 3}: $\sigma = c$. Recall there cannot be a 1 within $C$. As a prototypical example, suppose
$$r \quad =\quad  ( \cdots , a_1, a_4, 1, 2, \underbrace{c_2, a_2, 2,}_{C} \underbrace{c_1}_t, \underbrace{c_3, c_4, c_1,}_{C} c_1, c_2, a_3, c_3, \cdots ).$$
Similar to Subcase 2.2, the sets $X\in r$ that satisfy $t \ra \ol{C}$ are precisely the ones in the union of $r'$ and $r''$, where
$$\begin{array}{llcccccccccccccccccc}
r'& = & ( \cdots, & a_1, & a_4, & 1, & 2, & c_2, & a_2, & 2, & {\bf 0}, & c_3, & c_4, & c_1, & c_1, & c_2, & a_3, & c_3, & \cdots ), \\
\\
r'' & =& ( \cdots, & 0, & 2, & 1, & 2, & 0, & 0, & 0, & {\bf 1}, & 0, & 0, & 0, & 2, & 2, & a_3, & c_3, & \cdots ).
\end{array}$$
Indeed, setting $c_1 =1$ forces $a_1=0$ (whether or not $a_1$ happens to be in $C$); this in turn switches any potential $c_1$ outside $C$ to $2$. Similarly $a_2=0$ (being in $C$) forces $c_2=2$ outside $C$. Setting one $c_3$ in
$a_3 \ra \ol{c_3 c_3}$ to $0$ shortens the anti-implication to $a_3 \ra \ol{c_3}$, whereas setting $c_4$ in $a_4 \ra \ol{c_4}$ to $0$ yields $a_4 =2$.

The above shows that one is never forced to delete rows, and that imposing an anti-implication on a row of length $w$ costs $O(w)$. There are at most $R\le N\ (:=ms(G))$ rows $r_i$ in the final stack because they are mutually disjoint and each encodes at least one anticlique. Each $r_i$ arises by the enforcement of at most $w$ anti-implications (since some anti-implications may incidently hold without having been imposed). Since no deletions occur, no other (superfluous) work is done, and so the claimed  bound 
$O(Rw^2)=O(Nw^2)$ results. \hfill $\blacksquare$

From the proof it is clear that the number $R$ of final 012ac-rows satisfies 
$$(3)\quad R\ \le\  min(ms(G),2^{|Z|}),$$
where $Z$ is the vertex-cover (vc) on which  the (a,c)-algorithm was based. As seen,
 $R$ can be significantly smaller than $N=ms(G)$. On the other hand, when $N$ is small then $R\ (\le N)$ is even more so, and the chosen vertex-cover is rather irrelevant.
 An extreme example is the complete graph on $w$ vertices; then $N=w+1$ is tiny, yet $2^{|Z|}=2^{w-1}$ for all minimum vc $Z$. 
 
{\bf 2.6} Two more facts are noteworthy.
First,  the working stack (being a Last-In-First-Out stack) can at most contain $w$ rows at any given moment. 
Second, the {\it order} in which the quasi-implications are imposed  influences the magnitude of $R$. While no research on that issue exists, in later Sections we find other ways to improve the (a,c)-algorithm.

\section{A first look at pure Boolean 2-CNFs}

We assume a basic familiarity with Boolean functions, e.g. see [CH].
Recall that a {\it clause} is a disjunction of Boolean literals, such as $\ol{x}_1\vee x_3\vee \ol{x}_7$. A Boolean formula is in Conjunctive Normal Form (CNF) if it is a conjunction of clauses. If all clauses have at most two literals then on speaks of 2-CNFs. Each 2-CNF is readily reduced to an equivalent {\it pure} 2-CNF, i.e. one without 1-clauses.

{\bf 3.1} Let us couple a Boolean variable  $x_i$ to each 
vertex $i$ of $G_1$. Furthermore e.g. the edge $\{2,4\}$ becomes the  clause $\ol{x}_2\vee\ol{x}_4$. In this way the edge set of $G_1$ translates to this pure 2-CNF:
$$(4)\quad F(x):=(\ol{x}_1\vee\ol{x}_2)\wedge (\ol{x}_1\vee\ol{x}_4)\wedge (\ol{x}_1\vee\ol{x}_5)\ \wedge\ (\ol{x}_2\vee\ol{x}_4)\wedge(\ol{x}_3\vee\ol{x}_4)\wedge(\ol{x}_4\vee\ol{x}_5)\ \wedge\ (\ol{x}_5\vee\ol{x}_6)$$
Consider any  {\it model} $y$ of $F$, i.e. $y\in\{0,1\}^6$ makes $F(y)=1$. If some edge $\{i,j\}$ of $G_1$ was contained in $ones(y):=\{i:\ y_i=1\}$, one would have $\ol{y}_i\vee\ol{y}_j=0\vee 0=0$, which entails the contradiction $F(y)=(\ol{y}_1\vee\ol{y}_2)\wedge...\wedge 0\wedge...=0$. It follows that $ones(y)\in Acl(G_1)$. Conversely each anticlique $Y$ of $G_1$ is a model of $F$. Alternatively we say that $Y$ {\it satisfies} the 2-CNF formula\footnote{That's because setting $y_i:=1$ iff $i\in Y$ (and $y_i:=0$ otherwise), yields a bitstring $y\in\{0,1\}^6$ such that $F(y)=1$. We henceforth identify Boolean formulas with the Boolean functions $\{0,1\}^n\to\{0,1\}$ they induce.} $F$ in (4).
By  distributivity $(\ol{x}_1\vee\ol{x}_2)\wedge (\ol{x}_1\vee\ol{x}_4)\wedge (\ol{x}_1\vee\ol{x}_5)\equiv \ol{x}_1\vee(\ol{x}_2\wedge \ol{x}_4\wedge\ol{x}_5)\equiv x_1\ra(\ol{x}_2\wedge \ol{x}_4\wedge\ol{x}_5 )$. Here $\equiv$ signifies the {\it equivalence} of Boolean formulas, i.e they induce the {\it same} function $\{0,1\}^6\to\{0,1\}$. It hence follows from (4) that
$$(5)\quad F(x)\ \equiv\ [x_1\ra(\ol{x}_2\wedge \ol{x}_4\wedge \ol{x}_5)] \wedge\ [x_4\ra(\ol{x}_2\wedge \ol{x}_3\wedge \ol{x}_5)]\ \wedge\ [x_6\ra \ol{x}_5]$$
Whereas (4) consists of seven clauses, (5) consists of just three quasi\footnote{Quasi-implications (nonstandard terminology) should not be confused with {\it implications} which by definition are Boolean formulas of type $(x_h\wedge...\wedge x_i)\ra(x_j\wedge...\wedge x_k)$.
For instance, formula (7) is a conjunction of three implications.}-implications.
This idea, enhanced by the $(a,c)$-wildcard, lead to the $(a,c)$-algorithm. The remainder of Section 3 is only needed in Section 9 and (in order not to loose the story line) we recommend to skip it for the time being.

{\bf 3.2}  A 2-clause is {\it mixed} if it has one positive and one negative literal. Later on, in connection with graphs, we will encounter conjunctions of mixed 2-clauses such as
$$(6)\ f_4':=(\ol{y}_1\vee y_4)\wedge(\ol{y}_1\vee y_5)\wedge(\ol{y}_1\vee y_6)\wedge(\ol{y}_6\vee y_7)\wedge(\ol{y}_6\vee y_9)\wedge(\ol{y}_9\vee y_{10})\wedge(\ol{y}_9\vee y_{11}).$$
Similar to above distributivity triggers a conjunction of considerably fewer implications, i.e.
$$(7)\quad  f_4'\equiv\big(y_1\ra(y_4\wedge y_5\wedge y_6)\big)\ \wedge\ \big(y_6\ra(y_7\wedge y_9)\big) \ \wedge\ \big(y_9\ra(y_{10}\wedge y_{11})\big).$$
If dually to the $(a,c)$-wildcard we define the $(a,b)$-wildcard as $(a,b,b,...,b):=({\bf 0},2,2,...,2)\uplus ({\bf 1},1,1,...,1)$, then the $(a,b)$-algorithm of [W1] renders\footnote{Generally the $(a,b)$-algorithm is applicable whenever a family of implications with {\it singleton premises}, i.e. $ x_i\ra(x_j\wedge...\wedge x_k)$,  needs to be processed. } the modelset of $f_4'$ in (7) as the disjoint union of the 012ab-rows $\sigma_1,\sigma_2,\sigma_3$ in Table 5:

 \begin{tabular}{c|c|c|c|c|c|c|c|c|l} \hline
 & $y_1$ & $y_4$& $y_5$ & $y_6$ & $y_7$ & $y_9$ &$y_{10}$ &$y_{11}$& cardinality\\ \hline
 &   &  &   &   &  & & & \\ \hline
 $\sigma_1=$ & $a$ & $b$ & $b$ & $1$ & $1$ & $1$ & $1$ & $1$ &   $2^2+1$\\ \hline
 $\sigma_2=$ & $0$ & $2$ & $2$ & $0$ & $2$ & $0$ & $0$ & $2$ &   $2^4$\\ \hline
  $\sigma_3=$ & $0$ & $2$ & $2$ & $0$ & $2$ & $a$ & $1$ & $b$ &   $(2+1)\cdot 2^3$\\ \hline 
\end{tabular}

{\sl Table 5: Expressing $Mod(f_4')$ in terms of 012ab-rows}

For instance, this is why $\sigma_3\s Mod(f_4')$. Each bitstring $y=(y_1,y_4,...,y_{11})$ in $\sigma_3$ satisfies $y_1\ra(y_4\wedge y_5\wedge y_6)$ and $y_6\ra(y_7\wedge y_9)$   because of $y_1=0$ and $y_6=0$ respectively. Furthermore $y_{10}=1$   reduces  $y_9\ra(y_{10}\wedge y_{11})$ to $y_9\ra y_{11}$. The latter holds because of the wildcard $(a,b)$ in $\sigma_3$. 

{\bf 3.2.1} The $(a,b)$-algorithm was designed as a tool to compress the distributive lattice $D(P,\le)$ of all ideals of a poset $(P,\le)$. For instance, viewing the bipartite graph $B_3$ in Figure 9 of Section 9 as a height 1 poset, it holds that $|D(B_3,\le )|=1036$. The $(a,b)$-algorithm renders $D(B_3,\le)$ as disjoint union of sixteen 012ab-rows $r_i$, the "fattest" two of which comprising 300+180 ideals:

\begin{tabular}{c|c|c|c|c|c|c|c||c|c|c|c|c|c|c||c} 
 & 1 & 2& 3 & 4 & 5 & 6 & 7 & $\ol{1}$ & $\ol{2}$ & $\ol{3}$ & $\ol{4}$ & $\ol{5}$& $\ol{6}$ & $\ol{7}$& cardinality\\ \hline
 &   &  &   &   &   & &   &  &   &   &   &  & &\\ \hline
 $r_1:=$ & $b_1$ & $b_1$ & 2 & 2 & $b_3$ & $b_2$ & $b_2$ 
 & $a_1$ & $a_2$ &0 & 0 & $a_3$ & 0& 0& $(2+1)(4+1)^2\cdot 2^2=300$ \\ \hline
 $r_2:=$ & $b_1$ & $b_1$ & 2 & 2 & $b_3$ & 1 & $b_2$
 & $a_1$ & $a_2$ &0 & 0 & $a_3$ & 1& 0& $(2+1)^2(4+1)\cdot 2^2=180$ \\ \hline
 \end{tabular}

 {\sl Table 6: Expressing parts of  $D(B_3,\le)$  in terms of 012ab-rows}
 
The 012ab-rows in Table 5 and 6 will  resurface in Section 9. Until then we stick to 012ac-rows.

\section{About hypergraphs and the K\"onig-Egarv\'ary property}

This is a good place to briefly take a broader perspective. A {\it hypergraph} on a set $V$ is any family ${\cal H}={\cal H}_V\s {\cal P}(V)$. The sets $U\in {\cal H}$  are called {\it hyperedges}. Thus a graph is a hypergraph with all hyperedges of cardinality 2. 
In this case we speak of edges rather than hyperedges and keep on writing $G=(V,E)$ rather than $G=(V,{\cal H})$.
Most graph concepts generalize to hypergraphs.
For instance $X\s V$ is an {\it anticlique} of the hypergraph ${\cal H}_V$ if no hyperedge is contained in $ X$.
Further $Y\s V$ is a {\it hitting-set} of ${\cal H}:={\cal H}_V$ if $Y\cap U\neq \es$ for all hyperedges $U$. 
The  fact that
$$(\forall U\in {\cal H})\ U\not\s X\ \LRa\ (\forall U\in {\cal H})\ (V\setminus X)\cap U\neq \es$$
means that a subset of $V$ is an anticlique of $\cal H$ iff its complement is a hitting-set of $\cal H$. In this correspondence
the {\it maximal} anticliques map upon  the {\it minimal} hitting-sets. It is a notorious open problem whether the family $MHS({\cal H})$ of all minimal hitting-sets (dually: the family of all maximal anticliques) can\footnote{Observe however that from a practical point of view it is more important whether a trillion minimal hitting sets {\it can} be enumerated in compressed manner [W2].} be enumerated in output polynomial time [GV]. 

{\bf 4.1} If the hypergraph is a graph $G=(V,E)$ then the hitting-sets are the vertex-covers (vc) of $G$, and we get the {\it vc-acl duality}. As is well known, in the graph case all minimal vcs (and thus all maximal acls)  can indeed be enumerated in output polynomial time. 
Hence we will turn to greener pastures, i.e. to the enumeration of the family $MVC(G)$ of all minimum-cardinality vertex-covers (=mvcs), and dually to the family $MACL(G)$ (see (1)) of all maximum-cardinality anticliques (=macls). The common cardinality $\tau(G)$ of all $Z\in MVC(G)$ therefore satisfies
$$\tau(G)=|V|-\alpha(G)$$
and is hard to calculate. For bipartite graphs $G=B$ it evidently holds that $\tau(B)\le min(s_1,s_2)$, where $s_1,s_2$ are the sizes of the shores. The inequality can be sharp:

\begin{center}
\includegraphics[scale=1.1]{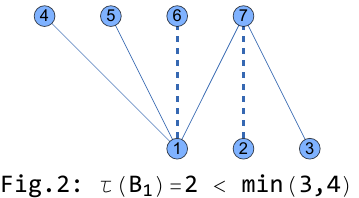}
\end{center}

{\bf 4.2} The hitting-sets of a hypergraph ${\cal H}_V$ are not only related to the anticliques of ${\cal H}_V$ but, equally important, also to the {\it matchings} $M$ of ${\cal H}_V$, i.e. to sets $M\s{\cal H}_V$ of {\it disjoint} hyperedges. It holds that
$$(8)\quad |M|\le |Z| \ for\ all\ matchings\ M\s{\cal H}_V\ and\ all\ hitting\ sets\ Z\s V\ of \ {\cal H}_V.$$
Indeed, if say $M=\{X_1,...,X_t\}$ then $\bigcup M=X_1\uplus\cdots\uplus X_t$, and so 
$$|M|=t\le |Z\cap(X_1\uplus\cdots\uplus X_t)|\le |Z|.$$ 
One says that a hypergraph is {\it K\"onig-Egarv\'ary (KE)} if for some matching $M$ and some hitting-set $Z$ it holds that $|M|=|Z|$.  In this case $M$ has necessarily maximum cardinality among all matchings, and $Z$ has minimum cardinality among all hitting-sets.
So far the KE property has been mainly investigated for graphs, and we will contribute to this endeavour in Sections 9 and 11.
Rephrasing (8) for graphs $G=(V,E)$ it reads:
$$(8')\quad |M|\le |Z| \ for\ all\ matchings\ M\s E\ and\ vertex\!-\!covers\ Z\s V\ of \ G$$
 Adopting standard notation let $\nu(G)$ be the maximum cardinality of a matching. Compared to $\tau(G),\alpha(G)$ one can calculate $\nu(G)$ fairly fast. That's because by Edmond's 1965 Theorem [S,p.418] a maximum matching can be found in time $O(w^2m)$, where $w:=|V|$ and $m:=|E|$.
  By $(8')$ it holds\footnote{The $\le$-gap can be extreme. For instance, in each $2w$-vertex complete graph $G$ one has $\nu(G)=w<2w-1=\tau(G)$.} that 
  $$\nu(G)\le \tau(G),\  and\ \nu(G)=\tau(G)\  takes\ place\ iff\ G\ is\ KE.$$
In particular, if $G$ is KE then $\alpha(G)$ and $\tau(G)$ can be calculated in time $O(w^2m)$.
For instance (check) both $G_1$ in Fig.1 and $B_1$ in Fig.2 are KE.

\section{A first look at $MACL(G)$ }

Subsection 5.1 presents a natural way to sieve $MACL(G)$ from the compressed representation of $Acl(G)$ derived in Section 2. In 5.2 we survey well known facts; namely $MACL(G)$ can be enumerated recursively in one-by-one fashion, and  for  perfect graphs one can achieve polynomial total time as well. Subsection 5.4 establishes, for any graph $G$, that $MACL(G)$ can be reduced to the calculation of $MACL(G[F^*])$ for a certain induced subgraph $G[F^*]$.

{\bf 5.1} For any 012ac-row $r$ we  put 
$$(9)\quad \alpha(r):=max\{|X|:\ X\in r\}\quad and\quad MACL(r):=\{Y\in r:\ |Y|=\alpha(r)\}.$$ 
(Do not confuse  $\alpha(r)$ and $\alpha(G)$.) So $r:=(a_1,c_1,c_1,\ a_2,c_2,c_2,\ a_3,c_3,c_3,c_3,\ 1,2,2,1,0)$ has $\alpha(r)=11$ and\footnote{Being pedantic (which we won't be henceforth) we point out that $MACL(r)$ is always a {\it set}, even when boiling down to a singleton like $\big\{\ (0,1,1,0,1,1,0,1,1,1,1,1,1,1,0)\ \big\}$.} $MACL(r)=(0,1,1,\ 0,1,1,\ 0,1,1,1,\ 1,1,1,1,0)$. As another example,\\
$MACL\big((a_1,c_1,\ a_2,c_2,\ a_3,c_3,\ 0,0,2,2)\big)=(a_1,c_1,\ a_2,c_2,\ a_3,c_3,\ 0,0,1,1).$
Generally, if $\rho$ has $t$ {\it short} wildcards $(a_i,c_i)$, i.e. of length two, then $|MACL(\rho)|=2^{t}$. 

For each graph  $G$ it holds that $\alpha(G)=max\{\alpha(r_i):\ 1\le i\le R\}$,
  where the $r_i$'s are the final rows produced by the (a,c)-algorithm, and so $Acl(G)=r_1\uplus\cdots\uplus r_R$.
  This provides one way  to find $\alpha(G)$ in the first place. Once $\alpha(G)$ is known, $MACL(G)$ can be sieved from the final rows $r_i$ in natural ways. Namely,
for the sake of notation let the indices $i$ that satisfy $\alpha(r_i)=\alpha(G)$ be $1,2,...,t\ (\le R)$. It then follows that $MACL(G)$ can be rendered as
 $$(10)\quad MACL(G)=MACL(r_1)\uplus\cdots\uplus MACL(r_t).$$
 To illustrate (10), a compression of $Acl(G_1)$ was obtained in Table 4. It follows from $\alpha(r_1)=\alpha(r_2)=\alpha(r_3)=3=\alpha(G)$ that
 $$(11)\ MACL(G_1)=MACL(\sigma_3)\uplus MACL(\sigma_4)\uplus MACL(\sigma_5)=\big\{\{1,3,6\},\{2,3,5\},\{2,3,6\}\big\}.$$
 Here $t=R$ but generally the gap $R-t$ is unpredictable. Therefore road (10) towards $MACL(G)$ (albeit crisp)  does not run in time polynomial in $t$ (contrasting Theorem 1 where we had $R$-linearity).

 Often combinatorial objects of maximum cardinality can be viewed as the macls (equivalently: maximum cliques) of some "auxiliary" graph. For instance, the maximum antichains of a poset $P$ are the macls of $P$'s  comparability graph. Or, the maximum matchings of any graph $G$ are the macl's in the line-graph of $G$.  Even if taylor-made algorithms exist for these and other problems, finding and carrying out these algorithms may take longer  (for small instances) than a calculation based on (10). In Section 12 we trim 012ac-rows to bip-val rows. Then idea (10), while still not output-polynomial, becomes a more serious competitor for calculating $MACL(G)$ (within a rather small crowd of competitors to begin with).

{\bf 5.2}  If $v\in V$ is fixed, then $MACL(G)$ splits into the classes $MACL(G,v)$ and\\ $MACL(G,\ol{v})$ of macls with and without $v$ respectively. Either one, but evidently not both, can be empty. No claim to originality is made for Lemma 2 (yet I can't find a reference). Adopting standard notation the {\it closed} neihborhood of vertex  $v$ is $NH[v]:=NH(\{v\})\cup\{v\}$.

{\bf Lemma 2:} {\it
\begin{enumerate}
\item[(a)] $MACL(G,\ol{v})\not=\emptyset$ if and only if $\alpha(\ G[V\setminus\{v\}]\ )=\alpha(G)$.
\item[(b)] $MACL(G,v)\not=\emptyset$ if and only if $\alpha(\ G[V\setminus NH[v]\ )=\alpha(G)-1$.
\end{enumerate} }

{\it Proof.} As to (a), both directions $\Ra$ and $\Leftarrow$ are obvious.

As to $\Ra$ in (b), any $X\in MACL(G,v)\not=\emptyset$ is an anticlique containing $v$, hence $X\cap NH[v]=\{v\}$, hence $X\setminus \{v\}$ is a cardinality $\alpha(G)-1$ anticlique of $G[V\setminus NH[v]\ ]$. It cannot be that $\alpha(G[V\setminus NH[v]])\ge\alpha(G)$ since otherwise $G$ had anticliques $Y\uplus\{v\}$ of cardinality $\ge \alpha(G)+1$, contradiction. Thus $\alpha(G\big[V\setminus NH[v]\big]\ )=\alpha(G)-1$.

 As to $\Leftarrow$ in (b), take any $Y\in Mmum(G[V\setminus NH[v]]\ )$, which hence has $|Y|=\alpha(G)-1$.
 Because of $Y\cap NH[v]=\es$ the set $X:=Y\cup\{v\}$ remains an anticlique. From $|X|=|Y|+1=\alpha(G)$ follows
$X\in MACL(G,v)$. Hence $ MACL(G,v)\not =\emptyset$. $\blacksquare$

Let $\cal K$ be a class of graphs which is closed under taking induced subgraphs.
Suppose there is an algorithm which for each graph $G\in {\cal K}$ with $w$ vertices produces a macl in time $\le T(w)$. Then a recursive search based on Lemma 2 shows that the $N$ macls of any graph $G\in{\cal K}$ can be enumerated in time $O(NwT(w))$. Trouble is, for the class $\cal K$ of all graphs the state of the art bound for $T(w)$ is $O(1.1996^w)$. 
If $\cal K$ is the class of all perfect graphs then  $T(w)=O(w^{6.5})$ seems to be (trusting AI) the current state of the art.
Because all comparability graphs of posets are perfect, one concludes that the $N$ maximum antichains of any $w$-element poset $(P,\le)$ can  be enumerated in time $O(Nw^{7.5})$. (This relates to the Dilworth lattice $D(P,\le)$ in Section 10.)
If $\cal K$ is the class of all bipartite graphs $B$ then $T(w)$ is $O(w^{0.5}m)$. That's because by [S,p.264] this is the time to find a maximum matching in $B$. In Section 11 we show that  $MACL(B)$ can actually be enumerated in a {\it compressed} manner, and that this extends to the class of all K\"onig-Egarv\'ary graphs (although the latter  no longer is a class of perfect graphs).

{\bf 5.3} If similarly to $MACL(G)$ in (1) we define $MaxAcl(G)$ as the family of all\\ (inclusion-)maximal anticliques of $G$, then\footnote{We mention in passing that a graph is called {\it well covered} [LM] if $MACL(G)=MaxAcl(G)$.}
$$(12)\quad MACL(G)\s  MaxAcl(G)\s Acl(G).$$
If $Acl(G)=r_1\uplus\cdots \uplus r_R$ then, contrasting (10), it only holds that
$$(10')\quad MaxAcl(G)\s Max(r_1)\uplus\cdots\uplus Max(r_R),$$
where by definition $Max(r_i)$ is the family of inclusion-maximal members of $r_i$. The fact that $4\in Max(\sigma_4)$ (see Table 4) but $\{4\}\not\in MaxAcl(G_1)$ proves that $\s$ in $(10')$ can be proper.
Many algorithms have been designed to find $MaxAcl(G)$  in output-polynomial  time. One of the first was the method  of Bron-Kerbosch (1973), aka Algorithm 457, which according Google Scholar has been cited 3761 times (as of 2026, sep-28). Recall from the Introduction, the article in front of you takes turns in tackling either $MACL(G)$ or $Acl(G)$.

 {\bf 5.4} By definition $core(G)$ is the intersection of all macls of $G=(V,E)$. For instance (11) implies that $core(G_1)=\{3\}$. Generally, having a type (10) representation of $MACL(G)$, and again assuming that $\{1\le i\le R:\ \alpha(r_i)=\alpha(G)\}$ is 
 $\{1,2,...,t\}$, it holds that 
  $$(\forall v\in V)\ (v\in core(G)\ \Leftrightarrow\  v\in ones(MACL(r_i))\ for\ all\ i\le t).$$
 This may look nice  but keep in mind that the mere decision $core(G)\stackrel{?}{=}\es$ is NP-hard. Putting aside calculating $core(G)$, Lemma 3 below shows how $core(G)$ "influences" the whole of $MACL(G)$. Some  preliminaries are in order.

 For each $U\s V$ the {\it neighborhood} $NH(U)$ consists of all $v\in U$ which are adjacent to some vertex in $U$. Put another way $NH(U)$ is the union of all sets $NH(v)$ when $v$ ranges over $V$.
 The shorthand $A^*:=core(G)$ insinuates\footnote{The superscript of $A^*$ will be explained in a moment.} that $A^*$ is the set of vertices appearing in {\it all} macls. Likewise $NO^*$  is the set of vertices appearing in {\it no} macl. Further $F^*:=MACL(G)\setminus (A^*\cup NO^*)$ is the set of all {\it flexible} vertices $v$ in that $v$ sits in some but not all macl's. Thus $V=A^*\uplus F^*\uplus NO^*$. Furthermore: 

$$(13)\quad NH(A^*)\s NO^*,\ hence\  NH(F^*)\s F^*\uplus NO^*\ and\ NH(NO^*)\s A^*\uplus F^*\uplus NO^*$$

The second and third inclusion in (13) are immediate consequences of the first. As to the first inclusion, $NH(A^*)\cap A^*=\es$ since otherwise there was an edge between members of $A^*$, which is impossible within an anticlique. Similarly, suppose there was an edge $\{u,v\}$ of $G$ with $u\in A^*$ and $v\in F^*$. By definition of $F^*$ there is a macl $X$ of $G$ with $v\in X$. But also $u\in X$ because $u$ is in {\it all} macls. This leads to the contradiction that the anticlique $X$ contains an edge $\{u,v\}$. This shows that besides
$NH(A^*)\cap A^*=\es$ also $NH(A^*)\cap F^*=\es$ takes place, and so $NH(A^*)\s NO^*$ in (13).
 For instance, in Figure 1 the vertices in $A^*,NO^*, F^*$ are colored white, black, blue respectively.

{\bf Lemma 3: }{\it For each graph $G$ it holds that 
$MACL(G)\stackrel{!}{=}\ {\cal S}:=\{Y\uplus A^*:\ Y\in MACL(G[F^*])\ \}$.}

{\it Proof.} Take any member $Y\uplus A^*$ of ${\cal S}$. It follows from $NH (A^*)\cap F^*=\es$ (see (13)) that $Y\uplus A^*$ is an anticlique in the first place. From $|Y\uplus A^*|=\alpha(G[F^*])+|A^*|$ follows $\alpha(G)\ge \alpha(G[F^*])+|A^*|$. Conversely,  pick $X\in MACL(G)$. From $V=A^*\uplus F^*\uplus NO^*$ and $X\cap NO^*=\es$ follows that $X=Y'\uplus A^*$ for some $Y'\in Acl(G[F^*])$. Hence $\alpha(G)=|X|=|Y'|+|A^*|\le \alpha(G[F^*])+|A^*|$. One concludes that
$\alpha(G)=\alpha(G[F^*])+|A^*|$ and that $MACL(G)={\cal S}$.
 $\square$

{\bf 5.5} Lemma 3 becomes the more relevant the more inviting the structure of $G[F^*]$ is. That comes to the fore in Section 11 where
 we  need a dual version of Lemma 3 which is in terms of mvcs and which we spell out now. Dually to before, for each graph $G=(V,E)$ it holds that $V=A\uplus F\uplus NO$, where by definition $A$ is the set of vertices appearing in {\it all} mvcs, $NO$ is the set of vertices appearing in {\it no} mvc, and $F:=V\setminus (A\uplus NO)$ is the set of all {\it flexible} vertices $v$ in that $v$ sits in some but not all mvcs. Furthermore
$$(14)\quad A^*=\{v\in V:\ (\forall X\in MACL(G))(v\in X)\}=\{v\in V:\ (\forall X\in MACL(G))(v\not\in X^c)\}$$
$$=\{v\in V:\ (\forall Y\in MVC(G))(v\not\in Y)\}=NO.$$
Similarly one shows that $NO^*=A$. Therefore $F^*=F$ and hence (see also Fig.3) Lemma 3 dualizes as follows:

{\bf Lemma 3': }{\it For each graph $G$ it holds that 
$MVC(G)=\{Z\uplus A:\ Z\in MACL(G[F])\ \}$.}

\begin{center}
\includegraphics[scale=0.4]{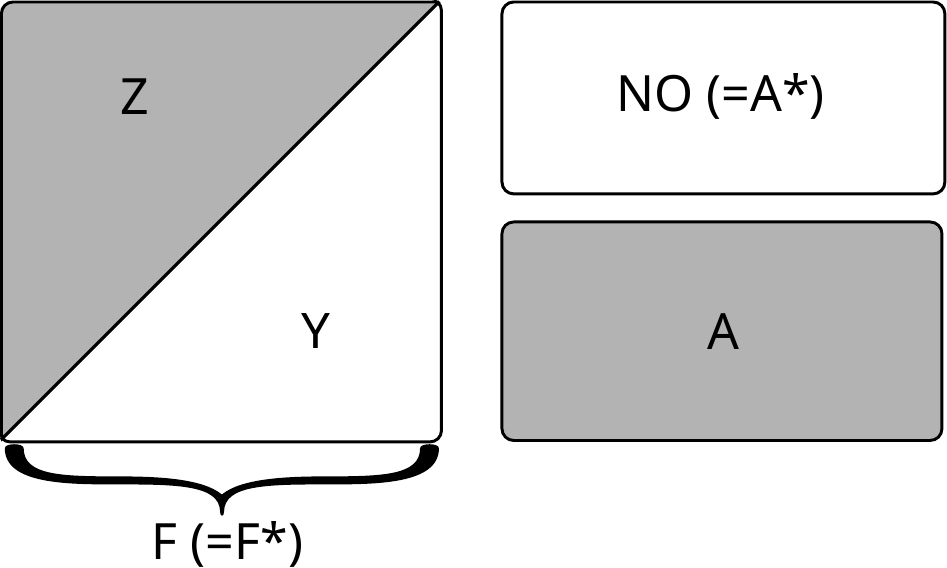}
\end{center}

{\tt Figure 3: The duality between Lemma 3 and Lemma 3'}

\section{Fixed cardinality anticliques}

The {\it independence polynomial}  of a graph $G$ is defined as
$$(15)\quad I(G,x) \quad : = 1+s_1x+s_2x^2+\cdots +s_kx^k+\cdots +s_\alpha x^\alpha\ (\alpha:=\alpha(G)).$$
Here $s_k$ is the number of $k$-element anticliques of $G$. Notice that $ms(G) = I(G, 1)$. 
The many applications of $I(G,x)$ (e.g. in statistical physics), and the many ways to compute $I(G,x)$, are surveyed in [LM]. While our way in 6.1 to calculate $I(G,x)$ presumably is not competitive because it requires a representation of $Acl(G)$ via 012ac-rows, it is lightning fast if this representation of $Acl(G)$ is available.
Sometimes the knowledge of all or some coefficients $s_k$ in (15) is not enough. What is desired (e.g. in [WYL] for the equivalent case of $k$-cliques) is an explicite knowledge of the family $Acl(G,k)$ of all $k$-element anticliques. In 6.2 we provide $Acl(G,k)$ in a compressed format.

{\bf 6.1} Each $s_k$ in (15) is the sum of all numbers
$$s_k(r) \quad : = \quad | \{X \in r: |X| =k \} |,$$
where $r$ ranges over all final 012ac-rows produced by the  (a,c)-algorithm.
Pleasantly $s_k(r)$ itself can be interpreted as the coefficient at $z^k$ of some {\it row-polynomial} pol$(r,z)$ in the variable $z$.  Namely, for each symbol $1$ in $r$ take a factor $z$, for each symbol 2 a factor $1+z$, for each short wildcard $ac$ a factor $1+2z$, and for each wildcard $ac\cdots c$ (with $m>1$ symbols $c$) a factor $1+(m+1)z + {m \choose 2} z^2 + {m \choose 3}z^3+ \cdots + {m \choose m} z^m$. That's because (say) $accc$ has 1 satisfying bitstring of weight 0, and $m+1=4$ of weight 1, and ${m\choose 2}=3$ of weight 2 (i.e. $0110,\ 0101,\ 0011$), and so forth.
It follows that the product pol$(r,z)$ of mentioned factors will do the job. For instance, if
$$ r_0 := (0,1,2,1,1,2,\ a_1, c_1, c_1, c_1,\  a_2, c_2,\ a_3, c_3, c_3, c_3, c_3)$$
then
$$\begin{array}{lll}
\mbox{pol}(r_0,z) & =& z^3\cdot(1+z)^2\cdot(1+4z+3z^2+z^3)\cdot(1+2z)\cdot(1+5z+6z^2+4z^3+z^4)\\
\\
& =& z^3+13z^4+70z^5+207z^6+379z^7+460z^8+383z^9+219z^{10}+83z^{11} + 19z^{12} + 2z^{13} \end{array}$$

The  highest term $2z^{13}$ signifies that $|MACL(r_0)|=2$. As to $MACL(r_0)$ itself, as argued in Section 5 we have
$$ MACL(r_0):=(0,1,1,1,1,1,\ 0,1,1,1,\ a_2,c_2,\ 0,1,1,1,1).$$

 {\bf 6.2} Suppose $Acl(G)$ is available as a disjoint union of 012ac-rows and we wish to sieve from it the family $Acl(G,10)$ of all 10-element anticliques. Proceeding row-wise represent each  012ac-row as a disjoint union of 012-rows in obvious ways. In Table 7 this is carried out for our 012ac-row $r_0$ (which is repeated on top of Table 7).

   \begin{tabular}{|c|c|c|c||c|l|} \hline
 $0\ 1\ 2\ 1\ 1\ 2$ & $a_1\ c_1c_1c_1$ & $a_2\ c_2$ & $a_3\ c_3c_3c_3c_3$ & applicable $k$'s \\ \hline
    &  &   &   &    \\ \hline
  $0\ 1\ 2\ 1\ 1\ 2$ & $0\ \ 2\ 2\ 2$ & $0\ 2$ & $0\ 2\ 2\ 2\ 2$ & $[3,\ 13]$ \\ \hline

    $0\ 1\ 2\ 1\ 1\ 2$ & $0\ 2\ 2\ 2$ & $0\ 2$ & $1\ 0\ 0\ 0\ 0$ & $[4,\ 10]$ \\ \hline

     $0\ 1\ 2\ 1\ 1\ 2$ & $0\ 2\ 2\ 2$ & $1\ 0$ & $0\ 2\ 2\ 2\ 2$ & $[4,\ 13]$ \\ \hline

     $0\ 1\ 2\ 1\ 1\ 2$ & $0\ 2\ 2\ 2$ & $1\ 0$ & $1\ 0\ 0\ 0\ 0$ & $[5,\ 10]$ \\ \hline

       $0\ 1\ 2\ 1\ 1\ 2$ & $1\ 0\ 0\ 0$ & $0\ 2$ & $0\ 2\ 2\ 2\ 2$ & $[4,\ 11]$ \\ \hline
        $0\ 1\ 2\ 1\ 1\ 2$ & $1\ 0\ 0\ 0$ & $0\ 2$ & $1\ 0\ 0\ 0\ 0$& $10\not\in[5,\ 8]$ \\ \hline
         $0\ 1\ 2\ 1\ 1\ 2$ & $1\ 0\ 0\ 0$ & $1\ 0$ & $0\ 2\ 2\ 2\ 2$ & $[5,\ 11]$ \\ \hline
          $0\ 1\ 2\ 1\ 1\ 2$ & $1\ 0\ 0\ 0$ & $1\ 0$ & $1\ 0\ 0\ 0\ 0$ & $ 10\not\in[6,\ 8]$ \\ \hline
\end{tabular}

{\sl Table 7:  Expanding $r_0$ into disjoint 012-rows}

For instance the second  row in Table 7, call it $\rho$,  contains $2^{10}=1024$ anticliques whose cardinalities range from 3 to 13.
Since there are ${10}\choose{7}$  ways to turn seven out of ten 2-bits of $\rho$ into 1-bits, there are ${{10}\choose{7}}=120$ anticliques of cardinality 10 in $\rho$. How to compactly express this set $S$ of 120 anticliques, viewing that one-by-one enumeration would be silly?

By definition\footnote{I used that tool before in other publications.} the wildcard $(g(t),g(t),...,g(t))$ means "exactly $t$ many 1-bits here"; say\\ $(g(3),g(3),g(3),g(3)):=\{(0,1,1,1),(1,0,1,1),(1,1,0,1),(1,1,1,0)\}$. 
It hence follows that
$$S=\big(0,1,g(7),1,1,g(7),\ 0,g(7),g(7),g(7),\ 0,g(7),\ 0,g(7),g(7),g(7),g(7)\big).$$
Similarly one sieves the 10-element anticliques from the other 012-rows in Table 7, except for those rows (the last and third last) which lack 10-element anticliques.

\section {The $(V_1,V_2)$-approach}

In this Section and the next we return from $MACL(G)$ to $Acl(G)$. Let $G=(V,E)$ and $V=V_1\uplus V_2$ be any bipartition. The $(V_1,V_2)$-approach  enhances the (a,c)-algorithm, and subsumes the latter as the special case $V_1=\es$ (Section 7). In Section 8 we fine-tune  "any bipartition" to "suitable bipartition".

{\bf 7.1} Suppose for some graph $G'$ on the vertex-set $V_2=\{1,2,...,w-1\}$ we have a representation $Acl(G')=r_1\uplus\cdots\uplus r_{t'}$ with $t'$ many
 012ac-rows $r_i$ of length $w-1$. In some  social (or other) applications one may need to {\it update} $G'$ because a new member $w$ joins his friends in the community $V_2$. Thus $V_2$ becomes $V:=\{w\}\uplus V_2$ and vertex $w$ is adjacent to some vertices in $V_2$. If $G=(V,E)$ is the new network, how can $Acl(G)$ be compressed with 012ac-rows of length $w$?  To recover $Acl(G')\s Acl(G)$, we attach a component $0$ at the end of each row $r_i$. 
 
 Concerning the arising {\it new} anticliques $X$ that contain $w$, consider say $NH(w)=\{7,12,18\}$. 
 Each row $r_i$ with $NH(w)\cap ones(r_i)\neq\es$ clearly must be canceled. Yet for
  each row $r_i$ with $NH(w)\cap ones(r_i)=\es$ attach a component $1$ at the end {\it and} set the 7th,12th and 18th component of $r_i$ to 0. This may force us changing other components of $r_i$ as well. The arising
 new 012ac-row, call it $(\{w\}\vee r_i)^\circ$, contains exactly those anticliques $X$ of $G$ that satisfy $w\in X$ and $(X\setminus\{w\})\in r_i$.

{\bf 7.2} Let us formalize and simultaneously generalize this matter. Thus for a given graph
$G=(V,E)$ we are asked to write $V=V_1\uplus V_2$ in such a manner that

\begin{itemize}
    \item[(16i)] $Acl(G[V_1])=\{X_1,X_2,...,X_t\}\quad(i\neq j\Ra X_i\neq X_j)$, and
    \item[(16ii)] $Acl(G[V_2])=r_1\uplus\cdots\uplus r_{t'}$ for  $t'$ many
    012ac-rows $r_i$.
\end{itemize}

We can assume\footnote{The case $V_1=\es$ can be interpreted as applying the plain (a,c)-algorithm to $G[V_2]=G$.} that $V_1\neq\es$ and that the empty anticlique $X_1:=\es$ is listed first in (16i). Thus $Acl(G[V_1])$ is rendered one-by-one, whereas $Acl(G[V_2])$ is compressed\footnote{Actually, it is compressed {\it unless} all $r_i$'s are 01-rows. But then the $(V_1,V_2)$-approach is pointless in the first place.}.
In view of Thm. 4 it is desirable to have both $t$ in (16i) and $t'$ in (16ii) as small as possible. (The scenario in 7.1 had $V_1=\{w\}=X_1$ and $t=1$.)

{\bf Theorem 4: } {\it With notation as in (16i),(16ii) it holds that $Acl(G)$ can be written as disjoint union of $s\le t\cdot t'$ many 012ac-rows.}

{\it Proof.} For each $D\in Acl(G)$ it holds that $D=(D\cap V_1)\uplus (D\cap V_2)=:X\uplus Y$. Obviously $X\in Acl(G[V_1])$ and $Y\in Acl(G[V_2])$.
Conversely, if $X\in Acl(G[V_1]),\ Y\in Acl(G[V_2])$, then $X\uplus Y$ needs {\it not} be an anticlique of $G$. It is one iff there are no edges connecting $X$ and $Y$. Putting
$$X_i\vee r_j:=\{X_i\uplus Y:\ Y\in r_j\}$$
it therefore holds that

$$Acl(G)\stackrel{!}{\s}\Big\{X\uplus Y: X\in Acl(G[V_1]),\ Y\in Acl(G[V_2])\Big\}=
\biguplus_{i\le t,\ j\le t'} X_i\vee r_j$$

The fact that some sets $X_i\uplus Y\s V$ are not anticliques will be mended (i.e. $\stackrel{!}{\s}$ becomes $=$)
by doing the following for each fixed pair $i,j$. {\bf Case 1:} $NH(X_i)\cap ones(r_j)\neq \emptyset$. Then we drop  $X_i\vee r_j$ because it contains no anticliques. {\bf Case 2:} 
$NH(X_i)\cap ones(r_j)= \emptyset$. Put $(X_i\vee r_j)^\circ:=(X_i\vee r_j)\cap Acl(G)$.
(In particular $(X_1\vee r_j)^\circ=(\es\vee r_j)^\circ=(\es\vee r_j)$.)
By inserting 0's in $r_j$ on the positions $k\in NH(X_i)\cap V_2$ and some other adaptions (best illustrated in Table 7) the set system $(X_i\vee r_j)^\circ$ can again be written as 012ac-row.  If $s$ is the number of pairs $i,j$ covered by Case 2, then $s\le t\cdot t'$.
  $\square$

{\bf 7.3} Let us give a concrete example where $t'=1$.  Recall from 2.2 the relevance of small vertex-covers (vc's). For each bipartite graph $B=(V,E)$ its\footnote{If $B$ is connected, then the shores are unique.} smaller shore $Z$ suggests itself as a vc of modest cardinality $|Z|\le\frac{1}{2}|V|$. For instance $B_2$ in Fig.4a  has $Z=\{1,2,..,6\}$ and upon applying the (a,c)-algorithm one would obtain (check) $Acl(B_2)$ as disjoint union of $R=12$
 (012ac)-rows. While $R=12$ already beats the a priori bounds $2^{|Z|}=64$ and\footnote{This calculation follows in a moment.} $ms(B_2)=556$ in (3), the $(V_1,V_2)$-approach will shrink $R=12$ to $R=5$.

{\bf 7.3.1} Here come the details. Puttig $V=V_1\uplus V_2$ with $V_1:=\{4,8,10\}$ and\\ $V_2:=\{7,1,2,\ 9,3,5,\ 6,11,12,13\}$, it holds that that 

\begin{itemize}
\item[]  $Acl(B_2[V_1])=\{X_1,...,X_5\}:=\{\es,\{4\},\{8\},\{10\},\{8,10\}\}$, and
\item[] $Acl(B_2[V_2])=(a_1,c_1,c_1,\ a_2,c_2,c_2,\ a_3,c_3,c_3,c_3)\ =:{\bf r}\quad  (indexed\ by\ 7,1,...,12,13)$.
\end{itemize}

\begin{center}
\includegraphics[scale=0.99]{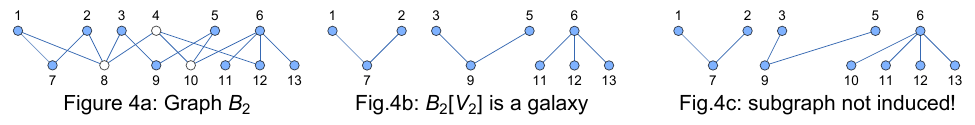}
\end{center}


According to Theorem 4 (and its proof) $Acl(B_2)$ is the disjoint union of the five 012ac-rows in Table 8. In particular $ms(B_2)=225+125+54+120+32=556$.

 \begin{tabular}{c|c|c|c||c|c|c||c|c|c||c|c|c|l|} \hline
 & 4 & 8& 10 & 7 & 1 & 2 & 9 & 3& 5 & 6 & 11 & 12& 13\\ \hline
 &   &  &   &   &   & &   &  &   &   &   & &\\ \hline
 $\rho_1:=\es\vee r=$ & $0$ & $0$ & $0$ & $a_1$ & $c_1$ & $c_1$ & $a_2$ & $c_2$ & $c_2$ &
 $a_3$ &  $c_3$ & $c_3$ & $c_3$ \\ \hline

  $\rho_2:=(\{4\}\vee r)^\circ=$ & $1$ & $0$ & $0$ & $a_1$ & $c_1$ & $c_1$ & $a_2$ & $c_2$ & $c_2$ &
 $a_3$ &  $c_3$ & $\bf 0$ & $c_3$ \\ \hline

  $\rho_3:=(\{8\}\vee r)^\circ=$ & $0$ & $1$ & $0$ & 2 &  {\bf 0} & {\bf 0} & $a_2$ &   {\bf 0} & $c_2$ &
 $a_3$ &  $c_3$ & $c_3$ & $c_3$ \\ \hline

  $\rho_4:=(\{10\}\vee r)^\circ=$ & $0$ & $0$ & $1$ & $a_1$ & $c_1$ & $c_1$ & $a_2$ & $c_2$ & $\bf 0$ &
 $\bf 0$ &  $2$ & $2$ & $2$ \\ \hline

  $\rho_5:=(\{8,10\}\vee r)^\circ=$ & $0$ & $1$ & $1$ & $2$ & $\bf 0$ & $\bf 0$ & $2$ & $\bf 0$ & $\bf 0$ &
 $\bf 0$ &  $2$ & $2$ & $2$ \\ \hline
 
\end{tabular} 

{\sl Table 8: Applying the $(V_1,V_2)$-approach to compress $Acl(B_2)$ }

We mention in passing that [CT] gives an efficient (recursive) formula for $ms(B)$ for each bipartite graph $B$.

{\bf 7.4} In order to explain how $V_2$ (and hence $V_1$) was found we need two handy notions. A {\it star} is a connected graph which possesses a vertex $a$ that lies on all edges. If there are no edges then the star is {\it trivial} and boils down to an isolated point $a$. Otherwise the star is {\it nontrivial}. If the star has at least two edges, $a$ is uniquely determined. Further, a {\it galaxy} is a graph $H$ all of whose connected components are stars. In this case $Acl(H)$ can be represented by some 2ac-row whose $2's$ match the isolated points of $H$ and whose (a,c)-wildcards match the nontrivial stars of $H$.

In view of Thm.4 it is desirable to find a subset $V_2\s V$ such that the induced graph $H:=G[V_2]$ is a galaxy.
This can be handled as follows. Given any graph $G=(V,E)$
let $Aux(G)$ be the auxiliary graph whose {\it vertex}-set is $E$, and where by definition two vertices $\{a,b\}$ and $\{c,d\}$ are adjacent iff 
either $G[\{a,b,c,d\}]$ is a triangle (and so $|\{a,b,c,d\}|=3$), or
$a,b,c,d$ (not necessarily in this order) are  the vertices of a length 3 path in $G$ (make a sketch). Consequently, if ${\cal A}\s E$ is an anticlique of $Aux(G)$ then no three members of $\cal A$ constitute a triangle or a length 3 path. Putting $V_2:=\bigcup{\cal A}$ it follows that $G[V_2]$ is a galaxy. 
 
 Which anticliques ${\cal A}$ of $Aux(G)$ are best for our purpose? Putting $V_1:=V\setminus V_2$ in view of 16(i) the value of $t:=ms(G[V_1])$ should be (in the best case) minimum. Let us argue that this forces $\cal A$ to be a maximal anticlique.
 Indeed, if there was an anticlique ${\cal A}':={\cal A}\uplus\{\{a,b\}\}$ in $Aux(G)$, then the edge $\{a,b\}$ of $G$ is not contained in $V_2$ (because the edge-set of the {\it induced} graph $G[V_2]$ is $\cal A$). If say\footnote{The case $\{a,b\}\cap V_2=\es$  is dealt with likewise.} $\{a,b\}\cap V_2=\{a\}$ then
  $V_2':=\bigcup{\cal A'}=V_2\uplus\{b\}$ and $V_1'=V_1\setminus\{b\}$, and so $ms(G[V_1'])< ms(G[V_1])$. While the maximal anticliques $\cal A$ of $Aux(G)$ can be enumerated in polynomial total time (see 5.3), scanning them all in order to get the minimum
  value  $ms(G[V_1])$ (where $V_1:=V\setminus\bigcup{\cal A}$) may take too long. Fortunately one can estimate the numbers $ms(G[V_1])$  accurately and fast [YLDCW]. It is therefore a sensible compromise to pick $V_2$ as the set for which the {\it estimate} of $ms(G[V_1])$ is minimum.

{\bf 7.5} Don't be confused by the following. The subgraph of $B_2$ in Fig.4c (with edge-set, say, ${\cal A^*}\s E$) looks like a galaxy even larger than the one in Fig.4b! Trouble is,  the subgraph induced by the vertices in $\bigcup {\cal A^*}$ has the additional edge $\{5,10\}$, and so this induced subgraph is {\it no} galaxy. What happened? Well, ${\cal A^*}$ is  no anticlique of $Aux(B_2)$; its vertices $\{9,5\}$ and $\{10,6\}$ are adjacent since $9,5,10,6$ are the vertices of a length 3 path in $B_2$.


\begin{center}
\includegraphics[scale=1.22]{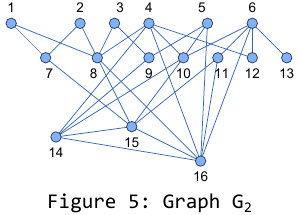}
\end{center}

{\bf 7.6}  Elaborating on 7.1, let us indicate how to {\it iterate} the $(V_1,V_2)$-approach.
Thus consider the (non-bipartite) graph $G_2=(V',E)$ in Fig. 5 and assume by induction that  a subset $\{1,2,...,13\}$ was found\footnote{How to find such a subset is the theme of Section 8.} that induces a bipartite subgraph $B$. That allows the standard (a,c)-algorithm (based on the vc $Z=\{1,...,6\}$) to achieve a good compression of $Acl(B)$. Actually $B$ is isomorphic to $B_2$ in Fig.4a. Recall that applying the $(V_1,V_2)$-approach to $B_2$ yielded the compression
of $Acl(B_2)=Acl(B)$ in Table 8. Let us apply the $(V_1,V_2)$-approach once more, i.e. to the new set $V_2:=\{1,...,13\}$, and hence $V_1:=V'\setminus V_2=\{14,15,16\}$. Since $V_1$ happens to be a 3-clique, $Acl(G[V_1])$ boils down to $\{\es,\{14\},\{15\},\{16\}\}$. The ensuing compression of $Acl(G_2)$ is rendered in the bottom half of Table 9.

 \begin{tabular}{c|c|c|c||c|c|c|c|c|c|c|c|c|c|c|c|l|} \hline
   & 14& 15 &16&  4 & 8& 10 & 7 & 1 & 2 & 9 & 3& 5 & 6 & 11 & 12& 13\\ \hline
 & &  &  &  & &  & &   & &   &  &   &   &   & &\\ \hline
$\es\vee\rho_1=$ & $0$ & $0$ &0& $0$ & $0$ & $0$ & $a_1$ & $c_1$ & $c_1$ & $a_2$ & $c_2$ & $c_2$ &
 $a_3$ &  $c_3$ & $c_3$ & $c_3$ \\ \hline
 $\es\vee\rho_2=$ & $0$ & $0$ & 0&    $1$ & $0$ & $0$ & $a_1$ & $c_1$ & $c_1$ & $a_2$ & $c_2$ & $c_2$ & $a_3$ &  $c_3$ & $0$ & $c_3$ \\ \hline
$\es\vee\rho_3=$ & $0$ & $0$ & 0&     $0$ & $1$ & $0$ & 2 &   0 &   0 & $a_2$ &   0 & $c_2$ & $a_3$ &  $c_3$ & $c_3$ & $c_3$ \\ \hline
$\es\vee\rho_4=$ & $0$ & $0$ & 0&  $0$ & $0$ & $1$ & $a_1$ & $c_1$ & $c_1$ & $a_2$ & $c_2$ & $0$ & $0$ &  $2$ & $2$ & $2$ \\ \hline
 $\es\vee\rho_5=$ & $0$ & $0$ &0&  $0$ & $1$ & $1$ & $2$ & $0$ & $0$ & $2$ & $0$ & $0$ &
 $0$ &  $2$ & $2$ & $2$ \\ \hline\hline

$(\{14\}\vee\rho_1)^\circ=$ & $1$ & $0$ &0& $\bf 0$ & $0$ & $\bf 0$ & $2$ & $0$ & $0$ & $\bf 0$ & $2$ & $2$ & $a_3$ &  $c_3$ & $c_3$ & $c_3$ \\ \hline
$(\{14\}\vee\rho_3)^\circ=$ & $1$ & $0$ &0& $\bf 0$ & $1$ & $\bf 0$ & $2$ & $0$ & $0$ & $\bf 0$ & $0$ & $2$ & $a_3$ &  $c_3$ & $c_3$ & $c_3$ \\ \hline

$(\{15\}\vee\rho_1)^\circ=$ & $0$ & $1$ &0& $0$ & $\bf 0$ & $\bf 0$ & $\bf 0$ & $2$ & $2$ & $a_2$ & $c_2$ & $c_2$ & $a_3$ &  $\bf 0$ & $c_3$ & $c_3$ \\ \hline
$(\{15\}\vee\rho_2)^\circ=$ & $0$ & $1$ &0& $1$ & $\bf 0$ & $\bf 0$ & $\bf 0$ & $2$ & $2$ & $a_2$ & $c_2$ & $c_2$ & $a_3$ &  $\bf 0$ & $0$ & $c_3$ \\ \hline

 $(\{16\}\vee\rho_1)^\circ=$     & $0$ & $0$ &1& $\bf 0$ & $\bf 0$ & $0$ & $a_1$ & $c_1$ & $c_1$ & $a_2$ & $c_2$ & $\bf 0$ &
 $\bf 0$ &  $\bf 0$ & $2$ & $2$ \\ \hline
 $(\{16\}\vee\rho_4)^\circ=$  & $0$ & $0$ & 1&  $\bf 0$ & $\bf 0$ & $1$ & $a_1$ & $c_1$ & $c_1$ & $a_2$ & $c_2$ & $\bf 0$ & $\bf 0$ &  $\bf 0$ & $2$ & $2$ \\ \hline\hline

\end{tabular} 

{\sl Table 9: Compressing $Acl(G_2)$ via $V=V_1\uplus V_2$, where $V_1$ is a clique and $G[V_2]$ is bipartite }

Generally in the $(V_1,V_2)$-approach it is best when $V_1$ is a clique (as in Table 9) because then $t$ in (16i) gets as small as $|V_1|+1$.
Conversely, $V_1$ being an anticlique seems to be  worst since then
$t=2^s$. But  looks can be deceiving: $V_1$ being an anticlique can have surprising perks as well.  More on that in Section 8.


\section{Feedback-vertex-sets and odd-cycle-covers }

 Given a graph $G=(V,E)$ we learned in 7.4 how to find $V_2\s V$ such that $G[V_2]$ is a galaxy. This is appealing because applying the $(V_1,V_2)$-approach (with $V_1:=V\setminus V_2$) to $G$ yields $t'=1$ in Theorem 4 and hence keeps $t\cdot t'$ moderate even for large values $t$. Trouble is, the denser $G$ the harder it is to find macroscopic galaxies at all. For instance, when $G$ is complete then all induced galaxies have cardinality at most two. Therefore, instead of being fixated to 
 a {\it special} type of forest (i.e. a galaxy), why not settle with an {\it arbitrary} forest $G[V_2]$?  As a perk this relates to long established research, i.e. $V_1:=V\setminus V_2$ is a so called feeback-vertex-set (8.1). Second, further weakening forests $G[V_2]$  to  bipartite graphs $G[V_2]$ leads to the well known concept (8.2) of an odd cycle cover $V_1$.
 
{\bf 8.1} A {\it feedback-vertex-set (fbvs)} of $G=(V,E)$ is a set $V_1\s V$ such that $G[V\setminus V_1]$ is a forest. Put another way, $V_1$ is a fbvs iff $V_1$ cuts all cycles of $G$. 
To fix ideas, one checks that $V_1:=\{1,3,10\}$ is a fbvs of $G_3$ in Fig.6a and that $G_4:=G_3[V\setminus V_1]$ is rendered correctly in Fig.6b. Putting $V_2:=V\setminus V_1$ let us carry out the $(V_1,V_2)$-approach on $G_3$.

 \begin{center}
\includegraphics[scale=1]{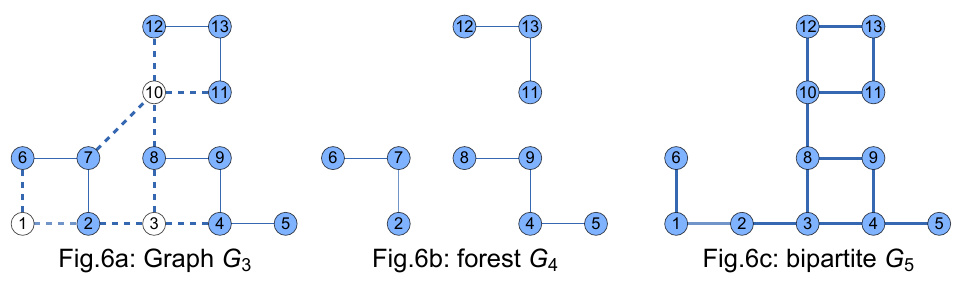}
\end{center}

One verifies that $r_1,r_2$ in Table 10  are such that $Acl([G_3[V_2])=r_1\uplus r_2$. Since $V_1$ happens to be an anticlique, the standard way (Thm.4) to proceed is to explicitely list all $t=2^3$ anticliques $X_i$ contained in $V_1$. There is a smarter way. As shown in Table 10, putting $X_8:=V_1=\{1,3,10\}$ it happens that $(X_8\vee r_j)^\circ\neq \es\ (j=1,2)$. A fortiori  $(X_i\vee r_j)^\circ\neq \es$ for all $1\le i\le 8$ and $1\le j\le 2$. This motivates the definition
$$r_1^*:=(X_1\vee r_1)^\circ\uplus (X_2\vee r_1)^\circ\uplus\cdots\uplus (X_8\vee r_1)^\circ\ \ (recall\ X_1:=\es\ by\ 7.2)$$
which can be captured (Table 10) by three {\it fancy don't-care} symbols $2^*$ on the positions that are indexed by the members of $V_1$ (i.e. 1,3,10). Likewise $r_2$ triggers $r_2^*$. In tandem the two {\it $0122^*ac$-rows} $r_1^*,r_2^*$ provide the hyper-compact representation
 $Acl(G_3)=r_1^*\uplus r_2^*$.

\begin{tabular}{c|c|c|c||c|c|c|c|c|c|c|c|c|l|} \hline
 & 1 & 3& 10 & 2 & 4 & 5 & 6 & 7& 8 & 9 & 11 & 12& 13\\ \hline
 &   &  &   &   &   & &   &  &   &   &   & &\\ \hline
 $r_1:=$ &  &  &  & $c_1$ & $a_3$ & $c_3$ & $c_1$ & $a_1$ & $2$ &
 $\bf 0$ &  $c_2$ & $c_2$ & $a_2$ \\ \hline

   $r_2:=$ & &  &  & $c_1$ & $0$ & $2$ & $c_1$ & $a_1$ & $0$ &
 $\bf 1$ &  $c_2$ & $c_2$ & $a_2$ \\ \hline

  &   &  &   &   &   & &   &  &   &   &   & &\\ \hline
 
 $(\es\vee r_1)^\circ=$ &0  &0  & 0 & $c_1$ & $a_3$ & $c_3$ & $c_1$ & $a_1$ & $2$ &
 $0$ &  $c_2$ & $c_2$ & $a_2$ \\ \hline
 $(\{1,3,10\}\vee r_1)^\circ=$ & $1$ & $1$ & $1$ & $0$ & $0$ & $2$ & $0$ & $0$ & $0$ & $0$ &  $0$ & $0$ & $2$ \\ \hline
  $r_1^*:=$ & $2^*$ & $2^*$ & $2^*$ & $c_1$ & $a_3$ & $c_3$ & $c_1$ & $a_1$ & $2$ &
 $0$ &  $c_2$ & $c_2$ & $a_2$ \\ \hline

  &   &  &   &   &   & &   &  &   &   &   & &\\ \hline

 $(\es\vee r_2)^\circ=$ &0 & 0 & 0 & $c_1$ & $0$ & $2$ & $c_1$ & $a_1$ & $0$ &
 $1$ &  $c_2$ & $c_2$ & $a_2$ \\ \hline
 $(\{1,3,10\}\vee r_2)^\circ=$ & $1$ & $1$ & $1$ & $0$ & $0$ & $2$ & $0$ & $0$ & $0$ & $1$ &  $0$ & $0$ & $2$ \\ \hline
 $r_2^*:=$ & $2^*$ & $2^*$ & $2^*$ & $c_1$ & $0$ & $2$ & $c_1$ & $a_1$ & $0$ &
 $1$ &  $c_2$ & $c_2$ & $a_2$ \\ \hline  
 
\end{tabular}

{\sl Table 10: Introducing the fancy don't care symbol $2^*$ to compress $Acl(G_3)$}

Admittedly, it requires extra work to know the anticliques in $r_1^*$ and $r_2^*$ more explicitely, e.g. just to count them. Nevertheless this much (which generalizes to 
arbitrary $0122^*ac$-rows) can be said:
$$(17)\quad 32=8\cdot\big|(V_1\vee r_1)^\circ\big|\ \le\  |r_1^*|\ \le\ 8\cdot\big|(\es\vee r_1)^\circ\big|=600$$

{\bf 8.2} Let us  hark back to the setting of Theorem 4 where the anticliques $X_1,...,X_t$ of $G[V_1]$ are listed one-by-one. If $X_\alpha$ and $X_\beta$ are such that $NH(X_\alpha)\cap V_2=NH(X_\beta)\cap V_2$ then $X_\alpha$ and $X_\beta$ cause the {\it same} 0's being inserted in each row $r_j$ (that occurs in Case 2 of Thm.4). This leads to an equivalence relation on the set $\{X_1,...,X_t\}$ such that each equivalence class $EC$ contains only $X_i$'s that  trigger the same 0's in each $r_j$. Trouble is, unless $V_1$ has exquisite properties, the members of a class $EC$ cannot (or only haphazard) be rendered in compressed format.

We now show that when $V_1$ is itself an anticlique then \footnote{Other kinds of exquisite $V_1$ are imaginable but remain to be discovered. Recall from 7.6 that $V_1$ being a clique does not make $V_1$ exquisite in terms of compressing $\{X_1,...,X_t\}$, but rather $t=|V_1|+1$ is so small that compression is pointless in the first place.} it is of this exquisite type. To begin with, we claim that for each fixed $r:=r_j$ there is a largest anticlique $U\s V_1$ such that $(U\vee r)^\circ\neq\es$.
Indeed,
$$U:=\bigcup\big\{\{x\}:\ x\in V_1\ and\ (\{x\}\vee r)^\circ\neq\es\big\}$$
does the job. (If $V_1$ is not an anticlique then $U$ needs not be an anticlique!) The set $U$ can be as small as $\es$ or as large as $V_1$, the latter e.g. taking place in Table 10.

{\bf 8.3} Some more theory needs to be digested before a toy example  can be formulated. Thus for any anticlique $Y\s U$ put $\ol{Y}:=\{x\in U:\ NH(x)\s NH(Y)\}$. Clearly $\ol{Y}$ is the largest superset of $Y$ that has the same neighbors as $Y$, and  $Y\mapsto \ol{Y}$  yields a closure operator on the powerset of $U$. As usual, a subset $Y\s U$ is called {\it closed} if $\ol{Y}=Y$. As is well known, the family $\cal L$ of all closed sets carries a lattice structure. More importantly, the equivalence classes $EC$ occuring in 8.2 now have a  neat structure. Specifically, there are as many $EC$'s as there are closed sets and each closed set $Y$ yields the class $EC(Y):=\{X\s U:\ \ol{X}=Y\}$. 

Two questions arise. First, how to calculate all closed sets efficiently? Second, how to compress the set system $EC(Y)$ for any given closed set $Y$?
As to the first question, given a closure operator $Y\mapsto \ol{Y}$ on some powerset ${\cal P}(U)$, various algorithms (e.g. the "cult" algorithm NextClosure which started it all)  have been proposed to calculate the lattice $\cal L$ of all closed sets, see [KO].
As to the second question, letting $K_1,...,K_p$ be the minimal members of $EC(Y)$ and writing $[K_i,Y]$ for the interval $\{X:\ K_i\s X\s Y\}$, it holds that $EC(Y)=[K_1,Y]\cup\cdots\cup [K_p,Y]$. (We refrain from discussing clever ways to calculate the $K_i$'s; instead see 8.5.)
Since each set system  $[K_i,Y]$ can be compressed as 12-row $\tau_i$ of length $|Y|$ (with $ones(\tau_i):=K_i$ and $twos(\tau_i):=Y\setminus K_i$) and since $EC(Y)=[K_1,Y]\cup\cdots\cup [K_p,Y]$ we conclude that
$$(18)\quad EC(Y)=\tau_1\cup\tau_2\cup\cdots\cup\tau_p.$$
While $(18)$ is better than listing $EC(Y)$ one-by-one, it would be nicer if the unions $\cup$ in $(18)$ could be replaced by disjoint unions $\uplus$ (upon altering the $\tau_i$'s to suitable $\sigma_i$). This will  be discussed in the upcoming toy example.

{\bf 8.4} Let us apply the $(V_1,V_2)$-approach to $G_6=(V,E)$ in Figure 7. 
Specifically,
$V_1:=\{1,...,6\}$ is a feedback-vertex-set (fbvs) which happens\footnote{How to get control over "happens" will be discussed in 8.6.} to be independent. Accordingly $V_2:=V\setminus V_1$ is a forest, incidently a galaxy, and so
 $Acl(G[V_2])=(c_1,a_1,c_1,c_2,a_2,c_2,c_2)=:r$ (indexed by 7,8,...,13) has cardinality 45. 

 \begin{center}
\includegraphics[scale=0.8]{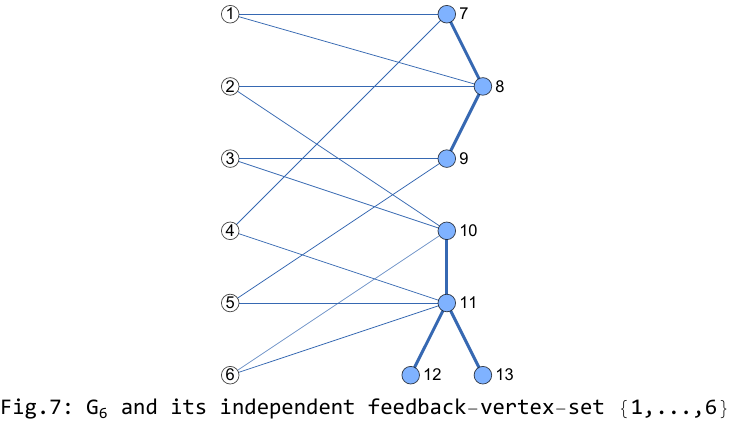}
\end{center}

In view of $ones(r)=\es$ we have $NH(X)\cap ones(r)=\es$ for all $X\s V_1$, which suffices for
the use of fancy don't-cares $2^*$ on the $V_1$-side. Thus $r^*$ in Table 11 renders the whole of $Acl(G_6)$, albeit momentarily  $$2^6\cdot 4\le |Acl(G_6)|\le2^6\cdot 45$$ 
is all  that can be said. This an instance of $(17)$ where $|(V_1\vee r)^\circ|=|\{\es,\{12\},\{13\},\{12,13\}\}|=4$
and $|(\es\vee r)^\circ|=|Acl(G[V_2])|=45$.

\begin{tabular}{c|c|c|c|c|c|c|||c|c|c|c|c|c|c|l|} \hline
 & 1 & 2& 3 & 4 & 5 & 6 & 7 & 8& 9 & 10 & 11 & 12& 13&   comment\\ \hline
 &   &  &   &   &   & &   &  &   &   &  & & &\\ \hline
 $r^*:=$ &$2^*$  &$2^*$  &$2^*$  & $2^*$ & $2^*$ & $2^*$ & $c_1$ & $a_1$ & $c_1$ &
 $c_2$ &  $a_2$ & $c_2$ & $c_2$ & $=Acl(G_6)$ \\ \hline

 &   &  &   &   &   & &   &  &   &  & &   & &\\ \hline
 $\rho_1:=$  &0  &0  &1  & 0 & 1 & 0 & $c_1$ & $a_1$ & 0 &
 0 &  0 & $2$ & $2$ & $|EC(\{3,5,6\})|=4$\\ \hline
$\rho_2:=$   &0  &0  &1  & 0 & 0 & 1 & $c_1$ & $a_1$ & 0 &
 0 &  0 & $2$ & $2$ & $\hspace{1cm} \vdots$\\ \hline
$\rho_3:=$   &0  &0  &{\bf 2}  & 0 & 1 & 1 & $c_1$ & $a_1$ & 0 &
 0 &  0 & $2$ & $2$ & $\hspace{1cm} \vdots$\\ \hline

 &   &  &   &   &   & &   &  &   &  & &   & &\\ \hline
$\sigma_1:=$  &1  &1  &2  & 2 & 1 & 2 & 0 & 0 & 0 &
 0 &  0 & $2$ & $2$ & $|EC(\{1,2,3,4,5,6\})|=26$\\ \hline
 $\sigma_2:=$  &1  &0  &1  & 1 & 2 & 2 & 0 & 0 & 0 &
 0 &  0 & $2$ & $2$ & $\hspace{2cm} \vdots$\\ \hline
  $\sigma_3:=$  &1  &1  &1  & 1 & 0 & 2 & 0 & 0 & 0 &
 0 &  0 & $2$ & $2$ & $\hspace{2cm} \vdots$\\ \hline
  $\sigma_4:=$  &1  &0  &1  & 0& 1 & 2 & 0 & 0 & 0 &
 0 &  0 & $2$ & $2$ & $\hspace{2cm} \vdots$\\ \hline
 $\sigma_5:=$  &1  &2  &1  & 0& 0 & 1 & 0 & 0 & 0 &
 0 &  0 & $2$ & $2$ & $\hspace{2cm} \vdots$\\ \hline
  $\sigma_6:=$  &1  &0  &0  & 2& 1 & 1 & 0 & 0 & 0 &
 0 &  0 & $2$ & $2$ & $\hspace{2cm} \vdots$\\ \hline
  $\sigma_7:=$  &0  &1  &1  & 1& 2 & 2 & 0 & 0 & 0 &
 0 &  0 & $2$ & $2$ & $\hspace{2cm} \vdots$\\ \hline
  $\sigma_8:=$  &0  &1  &0  & 1& 1 & 2 & 0 & 0 & 0 &
 0 &  0 & $2$ & $2$ & $\hspace{2cm} \vdots$\\ \hline
 \end{tabular}

 {\sl Table 11: The equivalence classes triggered by an independent fbvs $V_1$ (see Fig.7)}

 Referring to 8.3, it holds that $|{\cal L}|=20$; specifically,  there are $20=1+6+6+3+3+0+1$ closed sets $Y\s V_1$, having cardinalities varying between $0$ and $6$. Apart from the extremes $Y=\es$ and $Y=V_1$, e.g. the three 4-element closed sets $Y$ are $\{1,2,4,6\},\{2,3,5,6\},\{3,4,5,6\}$. As to the closed set $\{3,5,6\}$, its equivalence class is $EC(\{3,5,6\})=\big\{\{3,5\},\{3,6\},\{5,6\},\{3,5,6\}\big\}$. It follows that $\rho_1\uplus\rho_2\uplus\rho_3$ (Table 11) contains exactly those\footnote{Doing the same for the 19 other equivalence classes $EC(Y)$ yields 19 more numbers apart from 48, all of which summing up to $ms(G_6)$.} $12+12+24$ anticliques $S$ of $G_6$ that satisfy $S\cap V_1\in EC(\{3,5,6\})$.   The fat don't care 2 in $\rho_3$  manages to take care of $\{5,6\},\{3,5,6\}$ simultaneously.
In contrast to $EC(\{3,5,6\})$ the upcoming compression of the equivalence class $EC(V_1)$ will be more subtle. One checks ad hoc that the minimal members of $EC(V_1)$ are (shorthand notation) $125,134,135,136,156,234,245$. Consequently $EC(V_1)=(1,1,2,2,1,2)\cup(1,2,1,1,2,2)\cup\cdots\cup(2,1,2,1,1,2)$. Using techniques from [W3] one can rewrite this  as  {\it disjoint} union $EC(V_1)=\sigma_1'\uplus\cdots\uplus\sigma_8'$ where by definition $\sigma_i'$ is the 012-row consisting of the first six components of $\sigma_i$ in Table 11. For instance $(1,2,1,1,2,2)$, i.e. the interval $[\{1,3,4\},V_1]$, gets torn apart as $(1,{\bf 1},1,1,2,2)\uplus (1,{\bf 0},1,1,2,2)\s \sigma_1'\uplus\sigma_2'$.
More demanding is the interval $[\{1,3,6\},V_1]$; specifically the reader is invited to verify that
 $(1,2,1,2,2,1)\s \sigma'_1\uplus \sigma'_2\uplus \sigma'_3\uplus \sigma'_4\uplus \sigma'_5$.

{\bf 8.5} Given that $V_1$ is independent, calculating $\cal L$ and all the rest only makes sense  when $\cal L$ is significantly smaller than the powerset ${\cal P}(U)$. Setting a limit, say 100, one can launch one of the algorithms of [KO] in order to calculate $\cal L$ one-by-one. It either calculates some lattice $\cal L$ with $|{\cal L}|\le 100$, or the whole endeavour is aborted as soon the 101-th member of $\cal L$ is output. 
If $|{\cal L}|\le 100$, then the "whole endeavour" can be achieved brute-force\footnote{For $|U|\le 30$ this may well be feasible.} as follows. For all  $X\in {\cal P}(U)$ calculate $\ol{X}$. If $\ol{X}=X$ then $X\in {\cal L}$ has been encountered already. If $X\subset\ol{X}=:Y$ and $|X|=k$, add $X$ to a dynamically increasing
family $EC(Y,k)$ of all $k$-element members of $EC(Y)$. Having scanned ${\cal P}(U)$, one can calculate as follows for each $Y\in {\cal L}$ the family $Min[Y]:=\{K_1,..,K_p\}$ occuring in 8.3. Let $k_{min}$ be minimal with the property that $EC(Y,k_{min})\neq\es$. Evidently  $EC(Y,k_{min})\s Min[Y]$. Assume by induction that the "so far" family $Min'[Y]$ contains exactly the $\s$-inclusion minimal members of
$$EC(Y,k_{min})\uplus EC(Y,k_{min}+1)\uplus\cdots\uplus EC(Y,k).$$
Add exactly those $X\in EC(Y,k+1)$ to $Min'[Y]$ which have this property: There is no $X'\in Min'[Y]$ with $X'\s X$. Afterwards the so-far set $Min'[Y]$ has grown to a new so-far set $Min''[Y]$.

{\bf 8.6} An {\it odd-cycle-cover (occ)} $V_1\s V$ of a graph $G=(V,E)$ by definition makes $G[V\setminus V_1]$  bipartite. Equivalently, $V_1$ cuts all {\it odd} cycles of $G$. 
Consider again $G_3=(V,E)$ in Fig.6a. The odd cycle $C:=\{2,3,8,10,7\}$ prevents it from being bipartite. There are other odd cycles  but it is clear that any $u\in C$ yields an occ $\{u\}$, and so 
$G_3[V\setminus\{u\}]$ is bipartite. For instance, $G_5:=G_3[V\setminus\{7\}]$ is rendered in Figure 6c.

In most of  the standard applications of fbvs and occs it holds: the smaller they are the better.  Let $G=(V,E)=(\{1,...,n\},E)$ be any graph. Here comes [H] an elegant 0,1-program with $2n$ variables $c_v,s_v\in\{0,1\}$ that delivers a minimum-cardinality occ of $G$:

\begin{itemize}
    \item[] Minimize $c_1+\cdots +c_n$, subject to
  \item[(i)] $s_v+s_w+ (c_v+c_w)\ \ge 1\hspace{1.1cm} (\{v,w\}\in E)$
     \item[(ii)] $s_v+s_w- (c_v+c_w)\ \le 1\hspace{1.1cm} (\{v,w\}\in E)$
\end{itemize}

The explanation, in a nutshell, is as follows. Supposing that $V_1\s V$ is an occ of $G$, fix any $\{0,1\}$-coloring of the vertex-set $\{1,...,n\}\setminus V_1$ of the bipartite graph $G[V\setminus V_1]$.
Define $c_v:=1$ iff $v\in V_1$ (and 0 otherwise). Further, for $v\in V\setminus V_1$ define $s_v$ according to the color of $v$, and let $s_v\in\{0,1\}$ be arbitrary otherwise. Then both (i) and (ii) hold. (To verify this, for fixed edge $\{v,w\}$ make a case distinction according to the cardinality 2,1,0 of $\{v,w\}\cap V_1$.) The whole argument is reversible.

{\bf 8.6.1} But wait, doesn't $V_1$ being independent  matter more (for us) than the size $|V_1|$ ?
Yes indeed. Trouble is, it seems this quest can no longer be formulated as a 0,1-program because the condition
$$(iii) \quad \sum\Big\{c_vc_w:\ \{v,w\}\in E\Big\}\ =\ 0,$$
which forces $V_1$ to be independent, is not linear. But then again, (iii) can be handled by applying so called {\it Lagrange-multipliers}. 

{\bf 8.6.2}  If in contrast  we strive\footnote{Whether anticliques or cliques $V_1$ make more sense, depends of course on the density of $G$.} for $V_1$ being a large clique then one is led to replace the objective "Minimize $c_1+\cdots +c_n$" by
$$ Maximize\  \sum\Big\{c_vc_w:\ \{v,w\}\in E\Big\}.$$
While this does not guarantee  a clique $V_1$, this {\it quadratic} 0,1-program returns a dense enough  subgraph $G[V_1]$ which therefore has a moderate number $t$ of anticliques, as desired in the $(V_1,V_2)$-approach.


\section{Compressing $MVC(G)$ for KE graphs $G$\\ with a perfect matching}

In Sections 9 to 11 we return from $Acl(G)$ to $MACL(G)$.
 In Section 3 we  viewed $Acl(G)$ as the modelset of some 2-CNF  whose clauses have exclusively negative literals. The  (a,c)-algorithm  bundled these clauses in a way that invites wildcards.
 In a wholly different vein 2-CNFs also matter in the compression of $MVC(G)$ mentioned in the title. The underlying KE graphs with perfect matching are less haphazard than they seem. They pave the way to understanding $MACL(B)$ for bipartite graphs $G$.

{\bf 9.1}  Recall from Section 4 the duality between $MACL(G)$ and $MVC(G)$. In Section 9 it suits us better to stick to $MVC(G)$.
With each pure and satisfiable 2-CNF one can associate a certain "matched graph". Such graphs  are always K\"onig-Egarv\'ary  (KE) and furthermore possess a perfect matching. The converse is true as well.
 For instance $G_1$ (redrawn in Fig. 8A) is KE with perfect matching  $M_1:=\big\{\{1,2\},\{3,4\},\{5,6\}\big\}$. It hence is the matched graph of a certain 2-CNF (details later).

 \begin{center}
\includegraphics[scale=1.5]{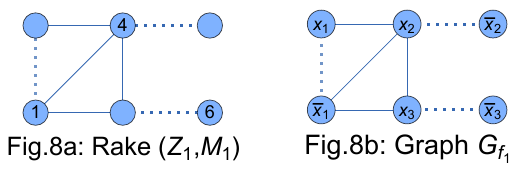}
\end{center}

Note that $G_1$ has a triangle and therefore is not bipartite. Nevertheless bipartite graphs will be our prime examples of KE graphs.

 {\bf 9.2} Let $G=(V,E)$ be any graph with a perfect matching $M\s E$. Then $\bigcup M=V$ and putting  $n:=|M|$ it follows that $|V|=2n$ and $\nu(G)=n$. If additionally $G=B$ is {\it bipartite} then each shore is a $n$-element vertex-cover. Hence $\tau(B)=\nu(B)$, i.e. $B$ is KE.
  The sparser $B$ the more likely there will be mvc's beyond the two shores, and the more likely $MVC(B)$ will in fact be compressible with $g$-wildcards.
 As an extreme  case, if $B$ has $n$ connected components, all of them single edges, then $MVC(B)=(g_1,g_1,g_2,g_2,...,g_n,g_n)$ has cardinality $2^n$. 
More interesting, take the connected bipartite graph $B_3=(V_3,E_3)$ in Figure 9, which has the perfect 
matching $M_3=\big\{\{1,\ol{1}\},...,\{7,\ol{7}\}\big\}\s E_3$. 

 \begin{center}
\includegraphics[scale=1.5]{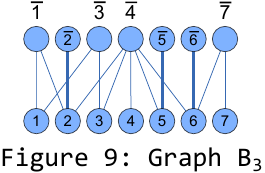}
\end{center}

 It turns out (details later) that $MVC(B_3)=r_1\uplus\cdots \uplus r_8$ has $2+4+\cdots+4=26$ elements (Table 12). For instance, the fact that $r_1\s MVC(B_3)$ can be argued as follows. The vertices $\ol{1},\ol{3},\ol{4},\ol{7}$ cover all edges of $B_3$ except the three fat ones. In each of the fat edges one can pick exactly one point, indicated by the three $g$-wildcards of $r_1$. Uniting  $\{\ol{1},\ol{3},\ol{4},\ol{7}\}$ with $2^3$ choices of 3-sets (e.g. $\{2,\ol{5},6\}$) yields 8 distinct mvc's $Z$ of $B_3$, and they (as bitstrings) constitute the members of $r_1$.

\begin{tabular}{c|c|c|c|c|c|c|c||c|c|c|c|c|c|c||c} 
 & 1 & 2& 3 & 4 & 5 & 6 & 7 & $\ol{1}$ & $\ol{2}$ & $\ol{3}$ & $\ol{4}$ & $\ol{5}$& $\ol{6}$ & $\ol{7}$\\ \hline
 &   &  &   &   &   & &   &  &   &   &   &  & &\\ \hline
 $r_1=$ & 0 & $g_1$ & 0 & 0 & $g_2$ & $g_3$ & 0     & 1 & $g_1$ &1 & 1 & $g_2$ & $g_3$ & 1&  8\\ \hline
  $r_2=$ & 1 & $g_1$ & 0 & 0 & $g_2$ & $g_3$ & 0     & 0 & 0 &1 & 1 & $g_2$ & $g_3$ & 1&  4\\ \hline
  $r_3=$ & 1 & 1 & 1 & 0 & $g_1$ & $g_2$ & 0     & 0 & 0 &0 & 1 & $g_1$ & $g_2$ & 1&  4\\ \hline
   $r_4=$ & 1 & 1 & 1 & 1 & 1 & 1 & 0     & 0 & 0 &0 & 0 & 0 & 0 & 1&  1\\ \hline

   $r_5=$ & 0 & $g_1$ & 0 & 0 & $g_2$ & 1 & 1     & 1 & $g_1$ &1 & 1 & $g_2$ & 1 & 1&  4\\ \hline 
   $r_6=$ & 1 & 1 & 0 & 0 & $g$ & 1 & 1     & 0 & 0 &1 & 1 & $g$ & 0 & 0&  2\\ \hline
   $r_7=$ & 1 & 1 & 1 & 0 & $g$ & 1 & 1     & 0 & 0 &0 & 1 & $g$ & 0 & 0&  2\\ \hline
    $r_8=$ & 1 & 1 & 1 & 1 & 1 & 1 & 1     & 0 & 0 &0 & 0 & 0 & 0 & 0&  1\\ \hline  
 \end{tabular}

 {\sl Table 12: Packing the 26 minimum vertex-covers of $B_3$ into 01g-rows.}

 {\bf 9.3} The following concept will be crucial. A  pair $(Z,M)$ is a {\it rake} of a graph $G$ if $Z$ is a vertex-cover of $G$, and $M$ is a matching of $G$, and every $v\in Z$ belongs to some edge of $M$, and every edge of $M$ is incident with exactly one vertex in $Z$. In formulas: $|Z|=|M|$ and $|\{v,v'\}\cap Z|=1$ for all $\{v,v'\}\in M$.
 In particular, if a graph has a rake then it must be a KE graph. More can be said [CH,p.223]:
 \begin{itemize}
   \item[(19)] {\it A graph is KE iff it has a rake.}
   \item[(20)] {\it A graph is KE iff for every maximum matching $M$ and every mvc $Z$ the pair $(Z,M)$ is a rake. }
 \end{itemize}
 
For instance, if $Z_1:=\{1,4,6\}$ then $(Z_1,M_1)$ is a rake of $G_1$ in 9.1, and $B_3$ in 9.2 has 26 rakes of type $(Z,M_3)$. If there is another perfect matching $M_3'$ of $B_3$ (is there?) then there are additional 26 rakes of type $(Z,M_3')$.

{\bf 9.4} K\"onig-Egarv\'ary graphs with perfect matchings (such as $G_1$ or $B_3$) can be generated systematically as follows. With each pure 2-CNF $f=f(x_1,...,x_n)$ we associate\footnote{It corresponds to the concept of [CH,p.210], except that we are dealing with 2-CNFs as opposed to 2-DNFs. Furthermore, our 2-CNFs being pure, no loops occur in $G_f$.} the so called {\it matched graph} $G_f$ whose $2n$ vertices are all literals $x_1,\ol{x}_1,...,x_n,\ol{x}_n$. As to the edges, on the one hand they correspond to the clauses of $f$, on the other hand also\footnote{Some  clauses $x_i\vee\ol{x}_i$ may be present in $f$ as well but, being tautological, they can  be dropped. It merely matters  that all {\it  edges} $\{x_i,\ol{x}_i\}$ be present.}
$\{x_1,\ol{x}_1\},...,\{x_n,\ol{x}_n\}$ are postulated as edges.
Consequently they constitute a perfect matching  $M_f$ of $G_f$.
For instance, if the pure 2-CNF is
$$(21)\quad f_1:=(x_1\vee x_2)\wedge (\ol{x_1}\vee x_2)\wedge (\ol{x_1}\vee x_3)\wedge (x_2\vee x_3)$$
then its matched graph $G_{f_1}$ is rendered in Fig.8(b). The modelset $Mod(f_1)$ can be determined by inspection.  If $z\in Mod(f_1)$ has $z_1=0$, then $z_2=1$ (because of $x_1\vee x_2$). Hence $z_3$ can be both (because of $(0\vee 1)\wedge(\ol{0}\vee 1)\wedge(\ol{0}\vee z_3)\wedge(1\vee z_3)=1$). But if $z_1=1$, then again $z_2=1$ (because of $\ol{x}_1\vee x_2$), and so $z_3=1$ (because of $\ol{x}_1\vee x_3$). To summarize
$$Mod(f_1)=\{(0,1,0),(0,1,1),(1,1,1)\}.$$
Notice that the corresponding "literal models", i.e. 
$$(22)\quad \{\ol{x}_1,x_2,\ol{x}_3\},\ \{\ol{x}_1,x_2,x_3\},\ \{x_1,x_2,x_3\},$$ all happen to be vc's of $G_{f_1}$. Because $G_{f_1}$ has a matching of cardinality 3, these vc's are in fact mvc's. 

{\bf 9.5} In order to generalize this in Theorem 5, we first generalize "literal model".
Thus, given $f:\{0,1\}^n\to\{0,1\}$, a {\it literal} model of $f$ is a set of $n$ literals which all evaluate to 1. While being a pun, let us embrace this name to capture an important concept that has been nameless so far. We write $LitMod(f)$ for the set of all literal models of $f$. Of course $LitMod(f)$ is bijective to $Mod(f)\ (\s \{0,1\}^n)$ in straightforward ways.

{\bf Theorem 5 (known\footnote{Again we adapt the material of [CH,p.223] to our terminology. }) }:
{\it  
\begin{itemize}
  \item[(a)] If $f$ is a pure and satisfiable 2-CNF, then $G_f$ is a KE graph with a perfect matching. Furthermore the literal models of $f$ coincide with the mvc's of $G_f$.
  \item[(b)] Each KE graph $H$ with a perfect matching is isomorphic to some matched graph $G_f$. 
\end{itemize}  }

{\it Proof.} As to (a), we previously stressed the crucial perfect matching of $G_f$.
But why is $G_f$ a KE graph? Because $f$  is satisfiable by assumption, models $z=(z_1,...,z_n)$ exist. If $Z$ is the corresponding literal model then by the satisfiability assumption $Z$ intersects all edges of $G_f$. So $Z$ is a vertex-cover. In fact
$(Z,M)$ is a rake of $G_f$ since $|Z\cap \{x_i,\ol{x}_i\}|=1$ for all $\{x_i,\ol{x}_i\}\in M$. Therefore $G_f$ is KE by (19). Since conversely each rake yields a unique literal model of $f$, there is a bijection between $LitMod(f)$ and $MVC(G_f)$.

As to (b), label the perfect matching $M$ of $H$ as $M=\big\{  \{x_1,\ol{x}_1\}, \ldots, \{x_n,\ol{x}_n\}\big\}$ and accordingly translate the edge-set of $H$ to a pure 2-CNF $f$. It follows that $H\simeq G_f$, {\it provided} $f$ is satisfiable. Why is $f$ satisfiable? Since $H$ is KE by assumption, by (19) any mvc $Z$ yields a rake $(Z,M)$.
Therefore for all $1\le i\le n$ exactly one literal in the edge  $\{x_i,\ol{x}_i\}$ belongs to $Z$, and at least one literal in every other edge of $H$ belongs to $Z$. Hence $Z$ is a literal model of $f$, and so $f$ is satisfiable.
 $\square$

Relabeling Figure 8a with literals yields Figure 8b which in turn triggers a pure 2-CNF that happens to coincide with $f_1$. In other words, $G_1\simeq G_{f_1}$. Accordingly the three mvc's of $ G_{f_1}$ in (22) translate to the
 mvcs\footnote{Their complements $\{2,3,5\},\{2,3,6\},\{1,3,6\}$ are therefore the macls of $G_1$, which coincides with the result in 5.1. As mentioned, for the time being we mainly stick to mvcs.} $\{1,4,6\},\{1,4,5\},\{2,4,5\}$ of $G_1$.

{\bf 9.6} Turning from $G_1$ to $B_3$, start again by labeling the perfect matching $M_3$ with literals. 
Adopting the natural way to do that leads to Figure 10a.
Letting $f_3$ be the pure 2-CNF triggered by this labeling it holds again (Thm.5) that $B_3\simeq G_{f_3}$.
In contrast to $G_1$ now all edges yield {\it mixed} clauses, i.e. consisting of one positive and one negative literal. So e.g. edge $\{\ol{x}_4,x_6\}$ yields the clause $\ol{x}_4\vee x_6$, which in turn can\footnote{Recall that for all Boolean functions $\ell,h$ the Boolean function $\ell\ra h$ is defined as $\ol{\ell}\vee h$.} be written as  $x_4\ra x_6$. 
Directing all edges of $G_{f_3}$ in this way (and dropping the edges in $M_3$) yields the
 digraph $D(f_3)$ in Fig.10b
Since $D(f_3)$ happens to have no directed cycles, it yields a poset $P(f_3)$ as usual in that the order relation $\ge$ is defined as
$$x_i\ge x_j\ \LRa\ i=j\ or\ there\ is\ a\ directed\ path\ x_i\ra x_k\ra\cdots\ra x_j.$$
Figure 10C is the ensuing poset diagram. Hence $u=(u_1,...,u_7)\in\{0,1\}^7$ is a model of $f_3$  iff $u_i\ra u_j$ evaluates to 1 for all arcs $x_i\ra x_j$ in Fig.10B. Because these arcs $u_i\ra u_j$ are never of type $1\ra 0$, a moment's thought shows that $u\in\{0,1\}^7$ is a model of $f_3$ iff $\{x_i: \ u_i=1\}$ is an (order) ideal of $P(f_3)$. 

{\bf 9.6.1} In each poset the set of its  minimal elements constitutes an  ideal $K^{min}$ with the perk that each subset of $K^{min}$ is an ideal as well. For the poset $P(f_3)$ in Fig.10C one has $K^{min}:=\{x_2,x_5,x_6\}$.  Written as literal model this  
is $\{\ol{x}_1,x_2,\ol{x}_3,\ol{x}_4,x_5,x_6,\ol{x}_7\}$. By Thm.5 it is a mvc of $B_3$. Indexing according to the ordering $x_1,...,x_7,\ol{x}_1,...,\ol{x}_7$ this mvc becomes\\
$Z:=(0,1,0,0,1,1,0;\ 1,0,1,1,0,0,1)$. In fact, $Z$ is one of eight  mvc's (corresponding to the subsets of $K_{min}$)
 comprised in the 01g-row $r_1$ of Table 12.
 Generally speaking one can, as for any 2-CNF, represent
  $Mod(f_3)$ as disjoint union of 012-rows. For instance $(0,2,0,0,2,2,0)\s Mod(f_3)$ because all 8 subsets of $K_{min}=\{x_2,x_5,x_6\}$ are ideals of $(P(f_3),\le)$. Since mvcs only arise upon stretching (ordinary) 01-models to double-length literal models, $(0,2,0,0,2,2,0)$ gets stretched
to $(0,g_1,0,0,g_2,g_3,0;\ 1,g_1,1,1,g_2,g_3,1)$, which equals $r_1$ in Table 12.

 Vice versa, let us e.g. look at $r_6$ in Table 12 and ask what ideals of $P(f_3)$ are encoded by the two members of $r_6$. One ideal is 
$\{x_1,x_2,x_6,x_7\}$ (check with Fig.10C that this is indeed an ideal). The other ideal is $\{x_1,x_2,x_6,x_7,x_5\}$.

\begin{center}
\includegraphics[scale=1]{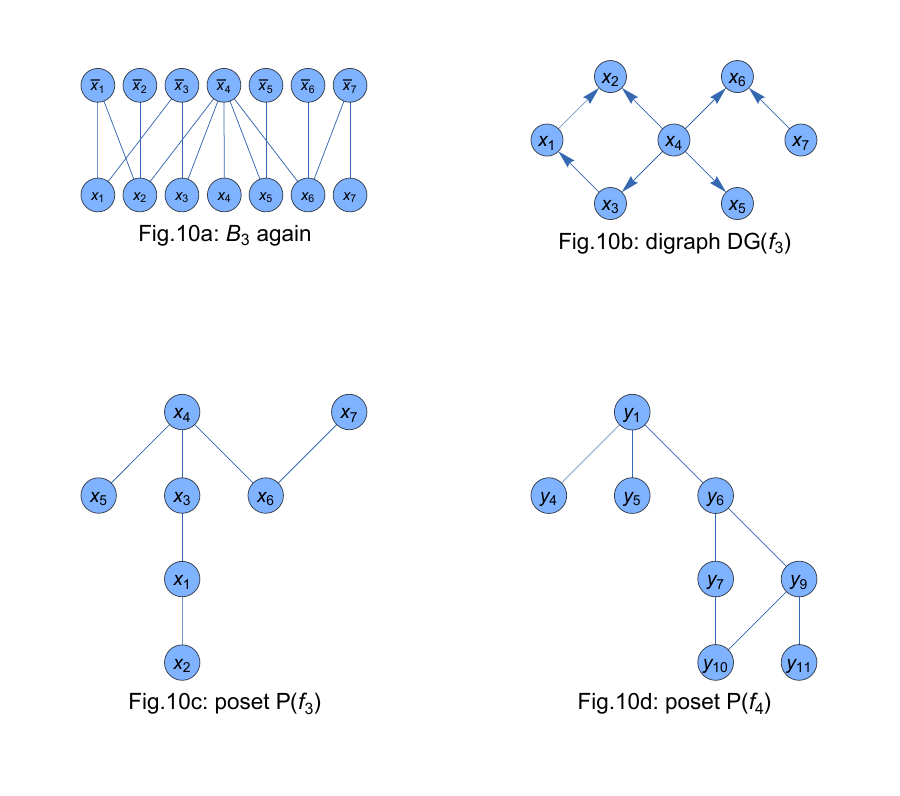}
\end{center}

{\bf 9.7} Let us tackle another, larger bipartite graph $B_4$ with perfect matching\\ $M_4=\big\{\{x_1,\ol{x}_1\},...,\{x_{11},\ol{x}_{11}\}\big\}$. Skipping its diagram, we directly head for
its characteristic digraph $DG(f_4)$, whose arcs again match the constituents $x_i\ra x_j$ of
$$ f_4:=(x_1\ra x_2)\wedge(x_1\ra x_2)\wedge(x_2\ra x_3)\wedge(x_3\ra x_1)\wedge(x_1\ra x_4)\wedge(x_3\ra x_5)\wedge(x_3\ra x_6)$$
$$\wedge\ (x_6\ra x_8)\wedge(x_6\ra x_9)\wedge(x_7\ra x_8)\wedge(x_8\ra x_7)\wedge(x_9\ra x_{10})\wedge(x_9\ra x_{11})$$
While $MVC(B_4)=LitMod(f_4)$ (Thm.5), let us focus on $Mod(f_4)$. As opposed to 9.6 we now confront two directed cycles in $DG(f_4)$, i.e. $x_1\ra x_2\ra x_3\ra x_1$ and $x_7\ra x_8\ra x_7$.
It helps that they  happen to coincide with the strong components in which they are contained\footnote{Recall that a strong component can contain {\it several} directed cycles.}. Having condensed the strong components $\{x_1,x_2, x_3\}$ and $\{x_7,x_8\}$ to $y_1$ and $y_7$ respectively, and upon identifying all other $x_i$ with $y_i$, the 2-CNF $f_4$ becomes a new (but much related) 2-CNF:
$$ f_4':=(y_1\ra y_4)\wedge(y_1\ra y_5)\wedge(y_1\ra y_6)\wedge(y_6\ra y_7)\wedge(y_6\ra y_9)\wedge(y_9\ra y_{10})\wedge(y_9\ra y_{11}).$$
Specifically, it holds that 
$$Mod(f_4)=\big\{(y_1,y_1,y_1,\ y_4,y_5,y_6,\ y_7,y_7,\ y_9,y_{10},y_{11}):\ (y_1,y_4,y_5,y_6,y_7,y_9,y_{10},y_{11})\in Mod(f_4')\big\}$$
It just so happens that $f_4'$ by distributivity coincides with (7) in 3.2. Therefore Table 5 compresses $Mod(f_4')$ by way of 012ab-rows; in particular this  yields $|Mod(f_4')|$ at once. While $$|Mod(f_4')|=|Mod(f_4)|=|LitMod(f_4)|=|MVC(B_4)|$$
(and this knowledge may sometimes be enough), it is a bit
cumbersome to carry over the 012ab-compression of $Mod(f_4')$ to a 01g-compression of $MVC(B_4)$ (as in Table 12). Suffice it to say that the first step is to expand each 012ab-row $r$ into disjoint 012-rows. (If $r$ has $t$ many $(a,b)$-wildcards then it triggers $2^t$ many 012-rows.)

\section{The Dilworth lattice of a poset}

The acls of a bipartite graph $B$ can be viewed as the antichains of a height 1 poset $(P,\le)$, and conversely. Hence our results about $MACL(B)$ promote the understanding of the Dilworth-lattice $Dil(P,\le)$ (at least for particular types of posets $(P,\le)$).

{\bf 10.1} Recall from 3.2.1 that $D(P,\le)$ is the distributive lattice of all ideals of a poset $(P,\le)$. As is well known, the maximal members of each ideal constitute an antichain, and conversely the "downset"
$X\!\downarrow\ :=\{y\in P:\ (\exists x\in X)\ y\le x\}$ of each antichain $X$ is an ideal. These correspondences yield mutually inverse lattice isomorphisms between $D(P,\le)$ and the lattice $Ach(P,\le)$ of all antichains of $(P,\le)$. In particular  the join-irreducible 
ideals $\{x\}\!\downarrow$ correspond to the singleton antichains $\{x\}$.

While the join (=supremum) and meet (=infimum) in $D(P,\le)$ is just (set-theoretic) union respectively intersection of ideals, the situation in $Ach(P,\le)$ is slightly more subtle. One checks that for all $X,Y\in Ach(P,\le)$ the join is $X\vee Y=Max(X\cup Y)$ (= set of $\le$-maximal members of $X\cup Y$), and the meet is
$X\wedge Y=Min(X\cup Y)$ (= set of $\le$-minimal members of $X\cup Y$). We write $P[X\cup Y]$ for the poset induced by $X\cup Y$ within $(P,\le)$. It is easy to see that each connected component of $P[X\cup Y]$ either has height 1 or is an isolated point. This in turn implies that
$$(23)\quad X\cup Y= (X\vee Y)\cup (X\wedge Y),\ \ X\cap Y\s (X\vee Y)\cap (X\wedge Y).$$
For instance, if $X:=\{1,2,3,4,5,6\}$ and $Y:=\{a,b,c,d,e,f,g\}$ are some antichains in a poset $(P,\le)$
such that $X\cap Y=\{5\}=\{g\}$, then $P[X\cup Y]$ may look as follows:

\begin{center}
\includegraphics[scale=0.8]{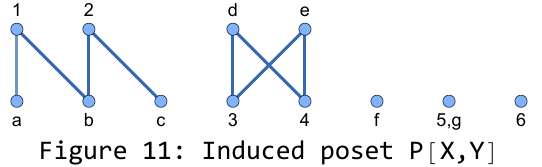}
\end{center}

Here $X\vee Y=\{1,2,5,6,d,e,f,g\},\  X\wedge Y=\{a,b,c,f,g,3,4,5,6\},\ (X\vee Y)\cap (X\wedge Y)=\{5,6,f,g\}.$
For later use we note the following consequence of (23):
$$(24)\ |X\wedge Y|+|X\vee Y|=|(X\wedge Y)\cup(X\vee Y)|+|(X\wedge Y)\cap(X\vee Y)|\stackrel{(23)}{\ge}
|X\cup Y|+|X\cap Y|=|X|+|Y|.$$

{\bf 10.2} Consider now the family $MACH(P,\le)\ \s \ Ach(P,\le)$ of all maximum-cardinality antichains. It turns out
that $MACH(P,\le)$ is in fact a sublattice of $Ach(P,\le)$, i.e.:
$$(25)\quad X\vee Y,\ X\wedge Y\in\ MACH(P,\le)\ for\ all\ X,Y\in MACH(P,\le)$$
{\it Proof of (25).} Streamlining previous work of Dilworth and Freese [F], the argument in [S,p.235] builds upon\footnote{The author liberated (23),(24) from being caught in between the lines of Schrijver's argument.}  (23),(24)  and goes as follows.
Because $X,Y$ have maximum cardinality one has $|X\wedge Y|\le |X|=|Y|$ and $|X\vee Y|\le |X|=|Y|$. In tandem with
$|X\wedge Y|+|X\vee Y|\stackrel{(24)}{\ge} |X|+|Y|$ follows that $|X\wedge Y|=|X\vee Y|=|X|=|Y|$. $\square$

{\bf 10.3} In 1979 Greene and Kleitman [GK] raised the problem of an inherent characterization of the join-irreducibles of $MACH(P,\le)$. Yet what "inherent" means, is debatable. For instance, given any member $X\in MACH(P,\le)$, it would be nice to have a polynomial time procedure deciding whether or not $X$ is join-irreducible. While the article [Fa] does not settle this particular question, it makes valuable contributions to the characterization issue.

A special role is played by posets $(B,\le)$ of height 1 since they can be identified with bipartite graphs $B=(V,E)$ in obvious ways. Vahagn Minasyan [M] claims\footnote{Minasyan never speaks of antichains. He defines a poset structure on $MACL(B)$ and then goes on to show the poset is a distributive lattice; being unaware of it being the Dilworth lattice.} to have an algorithm that constructs all join-irreducibles of $MACH(B,\le)$ in time $O(|V|\cdot |E|)$. I did not check the details, but if correct, Minasyan's algorithm still does not provide an inherent characterization of the join-irreducibles. 

{\bf 10.4} Let us connect the above to our results about bipartite graphs.
 As for any bipartite graph, the maximum anticliques of $B_3$ (Fig.9) coincide with the maximum antichains (=: machs) of the poset $(B_3,\le)$. Hence $MACH(B_3,\le)=MACL(B_3)$ and $|X|=7$ for all $X\in MACH(B_3,\le)$. It e.g. holds (check) that
$X':=\{x_1,x_2,x_3,x_4,\ol{x}_5,\ol{x}_6,\ol{x}_7\}$ and $X'':=\{\ol{x}_1,\ol{x}_2,x_3,x_4,\ol{x}_5,x_6,x_7\}$ are machs, and so $X'\vee X''=Max(X'\cup X'')=\{\ol{x}_1,\ol{x}_2,x_3,x_4,\ol{x}_5,\ol{x}_6,\ol{x}_7\}$ is a mach as well.
In 3.2.1 we mentioned that $|D(B_3,\le)|=1036$ and that $D(B_3,\le)$ can be compressed with 012ab-rows. As stated above, the Dilworth lattice $MACH(B_3,\le)$ is a sublattice of $Ach(B_3,\le)\simeq D(B_3,\le)$. 

How large is
$MACH(B_3,\le)$ and how can it (or its join-irreducibles) be captured? Its father $Ach(B_3,\le)$, compressed or not, is hardly helpful. Instead, as shown in 9.6, set up a certain pure 2-CNF $f_3$. It triggers a digraph $DG(f_3)$ (Fig.10b) which in turn triggers a poset $P(f_3)$ (Fig. 10c). It then holds that $MACH(B_3,\le)$ is isomorphic to $D(P(f_3),\le)$, which is shown in Fig.12.

{\bf 10.4.1} To fix ideas,
since $\{x_5,x_6,x_7\}$ is an ideal of $(P(f_3),\le)$, the corresponding literal model $\{\ol{x}_1,\ol{x}_2,\ol{x}_3,\ol{x}_4,x_5,x_6,x_7\}$ is a mvc of the graph $B_3$, and so its complement $\{x_1,x_2,x_3,x_4,\ol{x}_5,\ol{x}_6,\ol{x}_7\}$ is a macl of $B_3$, i.e. a member of $MACH(B_3,\le)$. In fact it is the mach $X'$ from above. Similarly the ideal $\{x_1,x_2,x_5 \}$ of $(P(f_3),\le)$ yields the mach
$X''$ from above. The union of the two ideals is the ideal $\{x_1,x_2,x_5,x_6,x_7 \}$ of $(P(f_3),\le)$. It triggers the mach $\{\ol{x}_1,\ol{x}_2,x_3,x_4,\ol{x}_5,\ol{x}_6,\ol{x}_7\}$, which coincides with $X'\vee X''$. This generalizes and thereby proves that the Dilworth-lattice $MACH(B_3,\le)$ is indeed isomorphic to $D(P(f_3),\le)$. In particular, $MACH(B_3,\le)$ has 26 elements and 7 join-irreducibles, corresponding to the 7 principal ideals $x_k\!\downarrow\ :=\{x_k\}\!\downarrow$ of $(P(f_3),\le)$). The  black dots in Figure 12 indicate the three members of $D(P(f_3),\le)$
(i.e. $x_5\!\downarrow\cup x_7\!\downarrow,\ \ x_1\!\downarrow\cup x_5\!\downarrow,\ \ 
x_1\!\downarrow\cup x_5\!\downarrow\cup  x_7\!\downarrow $) that map to $X',X'',X'\vee X''$ respectively.

\begin{center}
\includegraphics[scale=0.8]{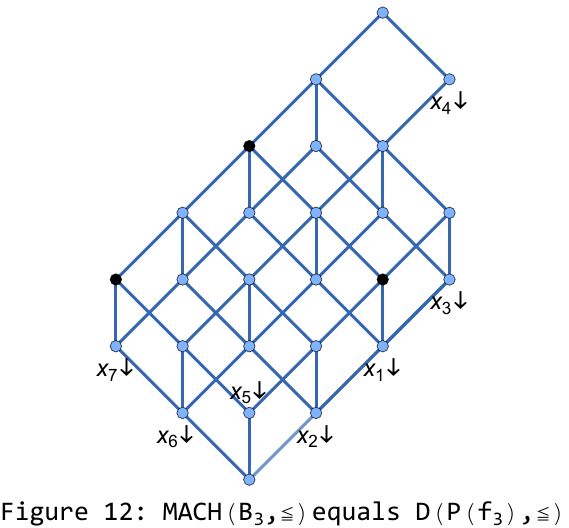}
\end{center}

{\bf 10.5} The facts about $B_3$ carry over to all bipartite graphs with perfect matching, and with little effort to arbitrary bipartite graphs (whose mvcs will be discussed in Section 11). How much of this can be adapted to posets of height $>1$, remains an open question. Recall from Subsection 5.2, if $(P,\le)$ is any poset then its comparability graph $Comp(P)$ is a perfect graph, and  so $MACH(P,\le)=MACL(Comp(P))$ can be enumerated in output-polynomial time. Since $MACH(P,\le)$ can be exponentially larger than the subset $J\big(MACH(P,\le)\big)$ of join-irreducibles, this leaves  open the 
question whether $J\big(MACH(P,\le)\big)$ can be generated in output-polynomial time as well.

\section{Compressing $MVC(G)$ for general KE graphs $G$ }

 After the Dilworth-lattice intermezzo we continue where we quit in Section 9. We now tackle K\"onig-Egarv\'ary (KE) graphs $G$ which {\it need not} have a perfect matching. Theorem 6 unravels the structure of $MVC(G)$ in this case. This leads to a output-polynomial (potentially compressed) enumeration of $MVC(G)$, and thus of $MACL(G)$. Special attention is paid to the most important subclass of KE graphs, i.e. the class of bipartite graphs.

{\bf 11.1} We previously observed that bipartite graphs with a perfect matching are evidently KE. Less\footnote{We recommend [S,p.263], which recasts an elegant argument of Ricci.} obvious, {\it each} bipartite graph is in fact KE. 
In Theorem 6 below everything depends on a fixed chosen maximum matching, which by 4.2 can be found in time $O(w^2m)$. 
 
\vspace{3mm}
{\bf Theorem 6: }{\it Let $G=(V,E)$ be a KE graph and $M$  a fixed maximum matching. 
\begin{itemize}
    \item[(a)]Putting $V_M:=\bigcup M$  the induced subgraph $H:=G[V_M]$ is
     KE with perfect matching $M$ and $MVC(G)\s MVC(H)$.
    
    \item[(b)] It holds that $A\s V_M$ and that the tail $T=T(G):=V\setminus V_M$ is an anticlique\\ contained in $ NO$.
   
    \item[(c)] If $NO_M$ is the set of siblings of $A$ w.r.t. $M$, then $NO=T\uplus NO_M$. 
 
    \item[(d)] It holds that $MVC(G)=\{A\uplus Z:\ Z\in MVC(G[F])\}$. Here the induced subgraph $G[F]$ is KE and has a perfect matching
    $M_F\s M$.
\end{itemize}
}

\begin{center}
\includegraphics[scale=0.4]{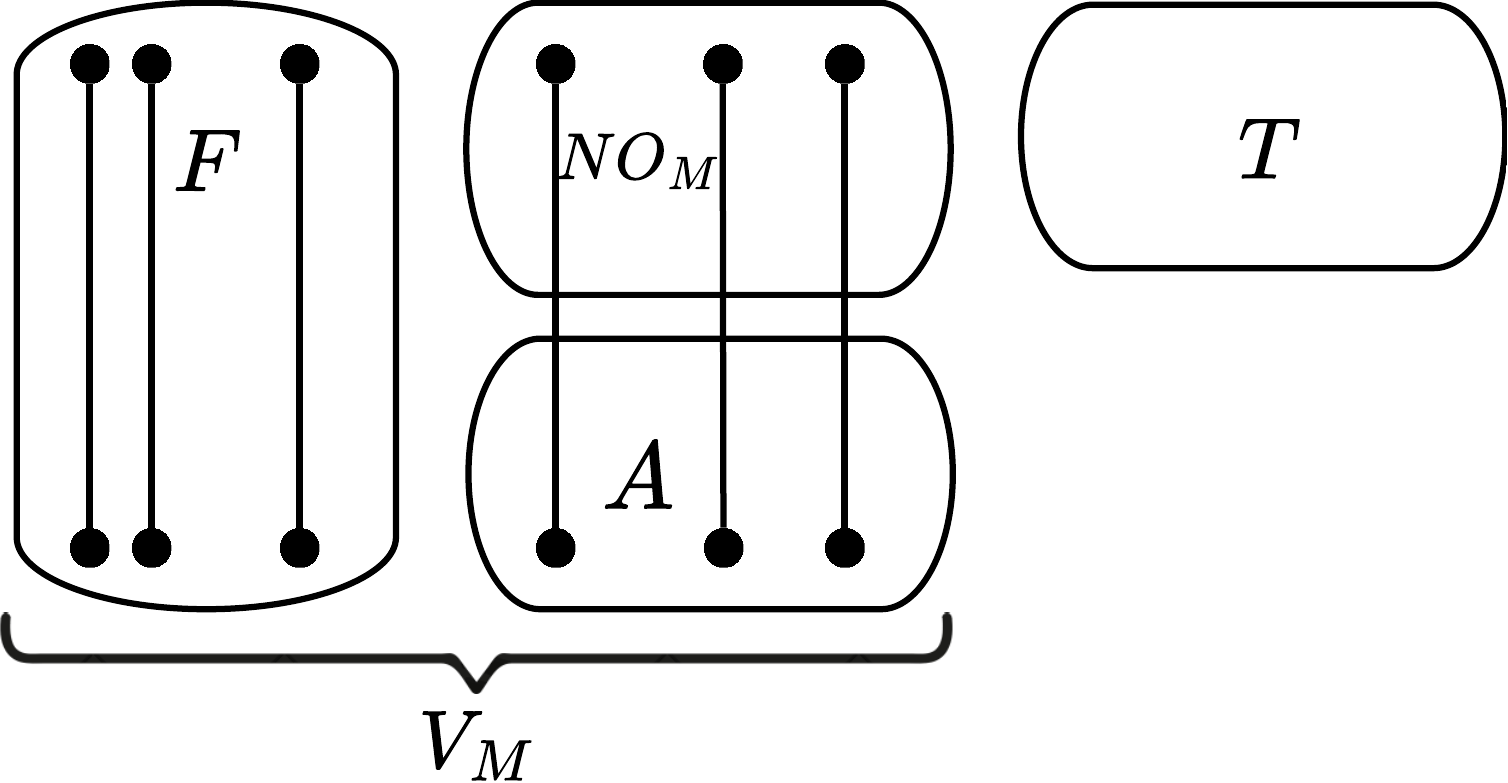}
\end{center}

{\tt Figure 13: The A-F-NO Theorem for K\"onig-Egarv\'ary  graphs}

As to Figure 13, recall from Section 5 the definition of $A,F,NO$. By duality statement (13) translates to:
$$(26)\ NH(NO)\s A,\ NH(F)\s A\uplus F,\ NH(A)\s A\uplus F\uplus NO$$
Furthermore $NO=NO_M\uplus T$ by Thm.6(c) and in Figure 13 only the edges of the maximum matching $M$ are indicated.  

\vspace{3mm}
{\it Proof.}  
As to (a), since $G$ is KE and $M$ is a maximum matching, each mvc  $Z$ of $G$ yields a rake $(Z,M)$ of $G$ by (20). A fortiori $(Z,M)$ is a rake of $G[V_M]$. Hence $H:=G[V_M]$  is KE by (19). Clearly $M$ is a perfect matching of $H$ and
 $MVC(G)\s MVC(H)$. 

As to (b), no vertex in the {\it tail} $T:=T(G)$ belongs to any mvc $Z$ of $G$ (since $Z\s V_M$ by (a)). Hence $T\s NO$. From $A\cap T\s
A\cap NO=\es$ follows that $A\s V_M$. If there were adjacent vertices $v,v'\in T$ then $M\cup\{\{v\}\}$ would be a matching larger than $M$. This contradiction shows that $T$ is an anticlique.

As to (c), each $v\in V_M$ has a unique {\it sibling} defined by $\{v,sib(v)\}\in M$. Since $(Z,M)$ is a rake for all $Z\in MVC(G)$, it holds for all $Z\in MVC(G)$ and all $v\in V_M$ that
$v\in Z\ \LRa\  sib(v)\not \in Z$.
In particular, it holds that 
$$(27)\quad (\forall v\in V_M)\quad v\in A\ \LRa\  sib(v) \in NO.$$ Hence if $NO_M:=\{sib(v'):\ v'\in A\}$
then $NO_M=NO\cap V_M$, and so $NO=T\uplus NO_M$.

As to (d), recall from Lemma 3' that $MVC(G)=\{Z\uplus A:\ Z\in MVC(G[F]) \}$ holds for {\it all} graphs $G$. As to the remaining claims,
clearly $M_F:=\big\{\{v,v'\}\in M:\{v,v'\}\s F\big\}$
is a perfect matching of $G[F]$ (see Fig.13). Take any $Z\in MVC(G)$. Each edge of $G$ contained in $F$ is touched not just by some (unpredictable) vertex $v\in Z$ but necessarily by some $v\in Z\cap F$. Hence $Z\cap F$ is a vertex-cover of $G[F]$, and from $|Z\cap F|=|M_F|$ follows that $G[F]$ is KE.
$\square$

\vspace{3mm}
{\bf 11.2}  Theorem 6 was partly inspired by [Z,Thm. 2] which for {\it trees} $G=(V,E)$ introduces the partition $V=A^*\uplus F^*\uplus NO^*$. The three statements in [Z,Thm.2] generalize from trees to KE graphs, or even arbitrary graphs. Specifically, two are covered by Theorem 6 and the third, translated from macls to mvcs, states that $deg(v)\ge 2$ for all $v\in A$.
Here's a short proof (facilitated by Fig.13). Take any $Z\in MVC(G)$ If $v':=sib(v)\in NO_M$ was the {\it only} neighbor of $v$ then $(Z\setminus\{v\})\cup\{v'\}$ would be a mvc as well, in contradiction to $v'\in NO$.

{\bf 11.3} According to Theorem 6(d) for any KE graph $G$ the calculation of $MVC(G)$ can be reduced to the calculation of $MVC(G[F])$, where $G[F]$ is a certain KE graph  which boasts a perfect matching. Hence $MVC(G[F])$ can be calculated via 2-CNFs by Theorem 5. Trouble is, in Theorem 6(d) one also needs to know $A=core(G)$! This (and more) is mitigated in the remainder of Section 11.

{\bf 11.4} Let $G=(V,E)$ be KE. If $M$ is a maximum matching and $V_M:=\bigcup M$ then the tail $T=V\setminus V_M$ is contained in $NO$ by Thm.6(c). The tail uniquely determines its {\it allmvc-shadow}, i.e. a certain set $A_S$ satisfying $A_S\s A:= A(G)$. For starters, $S_1:=NH(T)\s NH(NO)\s A$ by $(26)$. Hence $S_2:=sib(S_1)\s NO$ by $(27)$. Hence $S_3:=S_1\cup NH(S_2)\s A$ by $(26)$. Hence $S_4:=sib(S_3)\s NO$ by $(27)$. Hence
$S_5:=S_3\cup NH(S_4)\s A$ by $(26)$, and so forth.
The above process stabilizes as soon as $sib(S_{i+2})=sib(S_i)=:A_S$. Because $NO_S:=sib(A_S)\s NO$ we call $NO_S$ the {\it nomvc-shadow} of $T$. Putting  $$(28)\quad V':=V_M\setminus (NO_S\uplus A_S)$$
(which may be empty) it holds by construction of $NO_S$ that no edges go from $NO_S$ to $V'$, and also none from $T$ to $V'$. Furthermore by construction of $A_S$ one has $A_S\cap\{u,v\}\neq\es$ for all edges $\{u,v\}\in E$ that satisfy $u,v\in T\uplus NO_S\uplus A_S$. Put another way:

\begin{itemize}
    \item[(29)] {\it $A_S$ is a vertex-cover of $G[ T\uplus NO_S\uplus A_S]$ and the only potential edges between $NO_S\uplus A_S$ and $V'$ are between  $A_S$ and $V'$. }
\end{itemize}

{\bf 11.4.1} For illustration consider the KE graph $G_7=(V,E)$ (Fig.14) which has a maximum matching $M_7$ consisting of the eight boldface vertical edges. Hence the three leftmost vertices constitute the tail $T_7$ of $G_7$. They are labeled with $"0"$ to stress the fact that they belong to {\it no} mvc of $G_7$. Likewise vertices that belong to {\it each} mvc will carry labels $"1"$ (with subscripts). Adopting the above notation the set $S_1=NH(T_7)$ contains the two vertices labeled $"1_1"$. Because set notation implies that $\{1_1,1_1\}=\{1_1\}$, we rather write $S_1=[1_1,1_1]$. Hence
$S_2=[0_2,0_2]$ and $S_3=[1_1,1_1,1_3,1_3]$
(one vertex $1_3$ is triggered by two $0_2$'s). Further $S_4=[0_2,0_2,0_4,0_4],\ S_5=[1_1,1_1,1_3,1_3,1_5],\ S_6=
[0_2,0_2,0_4,0_4,0_5],\ S_7=S_5=A_S$. The three dashed edges in Fig.14 are the ones that lead from $A_S$ to $V'$.
As stated in $(29)$, check that all neighbors of 0-labeled vertices are 1-labeled, and so $A_S$ is a vertex-cover of $G[T\uplus NO_S\uplus A_S]$. We mention in passing that the six rightmost vertices in Fig.14 induce a subgraph isomorphic to $G_1$.

\begin{center}
\includegraphics[scale=1.3]{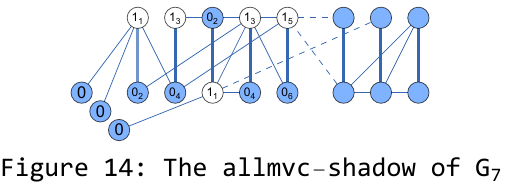}
\end{center}

{\bf 11.5} Observe that Thm.7(i) below resembles Thm.6(d) in that both $G[F]$ and $G[V']$ are KE graphs with a perfect matching. Crucially, as opposed to $A(G)$, the allmvc-shadow $A_S$ can be calculated fast.

{\bf Theorem 7: }{\it Let $G=(V,E)\ (w:=|V|,m:=|E|)$ be a KE graph with allmvc-shadow  $A_S$ (induced by the tail $T(G)$). Then the following holds:
 \begin{itemize}
\item[(i)] $MVC[G]=\big\{A_S\uplus Z':\ Z'\in MVC(G[V'])\ \big\}$. Here $V'$ is defined in $(28)$ and is such that $G[V']$ is KE with a perfect matching.
\item[(ii)] The $N$ members of $MVC(G)$ can be calculated in time $O(w^2m+Nw^3)$.
\item[(iii)]  $A(G)=A_S\uplus A(G[V'])$.
\end{itemize}
}

{\it Proof.} As to (i), analogous to $G[F]$ in Thm.6(d) one shows that $G[V']$ is KE and has a perfect matching $M'\s M$ (so $M'$ contains those edges of $M$ which are disjoint from $A_S$). In order to
 show $\s$ in (i), take $Z\in MVC(G)$. Then $Z=A_S\uplus (Z\cap V')$ because of $Z\s V_M$ and $(28)$. 
 Furthermore $Z':=Z\cap V'$ belongs to $MVC(G[V'])$ because $|Z'|=|M'|$. As to $\supseteq$, let $Z'\in MVC(G[V'])$. Then $A_S\uplus Z'$ is a vertex-cover of $G$ by $(29)$, and a minimum one because $|A_S\uplus Z'|=|M|$. 
 
 As to (ii), find a maximum matching $M$ of $G$ in time $O(w^2m)$ and calculate the two shadows $A_S,NO_S$ in even less time. Since $V'=V\setminus (A_S\uplus NO_S)$ and
 since $G[V']$ is KE and has a perfect matching $M'$, there is by Thm.5(b) a pure 2-CNF $f$ such that $G[V']\simeq G_f$ and  $MVC(G_f)=LitMod(f)$. By the definition of $N$ therefore $N=|MVC(G)|=|Mod(f)|$.
Since according to [W4, Thm.1] one can enumerate $Mod(f)$ in time $O(Nw^3)$, it follows that $MVC(G)$ can be enumerated in time $O(w^2m+Nw^3)$.

 As to (iii), this follows at once from (i).
 $\square$

{\bf 11.6} Recall from the proof of Thm.7(ii) that $MVC(G[V'])$ is bijective to $LitMod(f)$ for some
Boolean function $f:\{0,1\}^n\to\{0,1\}$ in 2-CNF format. The mentioned enumeration [W4] of $Mod(f)$ renders $Mod(f)$  as disjoint union of 012-rows. Each of them must however be morphed to a double-length 01g-row in order to properly encode a set of mvcs of $G[V']$. (To refresh memory, $(0,2,0,0,2,2,0)$ in 9.6.1 triggered the 01g-row $r_1$ in Table 12.)
According to Thm.7(i) each $01g$-row contributing to $MVC(G[V'])$ must be extended with $|A_S|$ many 1-bits. The collection of these extended $01g$-rows compresses $MVC(G)$.

{\bf 11.6.1} For bipartite graphs (which constitute a subclass of KE graphs) we observed in 9.7 that all clauses of $f$ are mixed. Consequently $Mod(f)$ could be rendered via 012ab-rows. Are they better than 012-rows? Yes and no. Yes, because 012ab-rows tend to be fewer\footnote{If $r$ has $s$ many (a,b)-wildcards then $r$ can be written as disjoint union of $2^s$ many 012-rows.}, which e.g. has the advantage that the cardinality $|MVC(G)|=|Mod(f)|$ can be calculated very fast. But also no, because in order to compress $MVC(G[V'])$ via 01g-rows one must first replace all 012ab-rows $r$ by 012-rows.

{\bf 11.6.2} From the proof of Theorem 7(i) it is clear that it can be generalized as follows. Defining the allmvc shadow, as well as $V'$, makes sense for an {\it arbitrary} graph $G=(V,E)$ as soon as a maximum matching $M\s E$ is chosen. It then holds that $G$ is KE iff $G[V']$ is KE. Since $G[V']$ has a perfect matching, it is KE iff a certain 2-CNF is satisfiable (Thm.5(i)).

{\bf 11.7} Let us translate Theorem 7(i) from a mvc-framework to a macl-framework. We begin by recalling old terminology and introducing new one (in all four cases "shadow" means "shadow of the tail $T(G)$") :
\begin{itemize}
    \item {\it allmacl-shadow} $A_S^*:=NO_S$ (= our previous nomvc shadow)
    \item {\it nomacl-shadow} $NO_S^*:=A_S$ (= our previous allmvc shadow)
\end{itemize}

{\bf Theorem 8: } {\it For each KE graph $G=(V,E)$ with fixed maximum matching $M\s E$ (and the ensuing parameters defined above) it holds that $$MACL(G)=\big\{T(G)\uplus A_S^*(G)\uplus X':\ X'\in MACL(G[V'])\big\},$$
where $G[V']$ is an induced KE graph with a perfect matching.}

{\it Proof.} It follows from $(28)$ and the above that
$$(30)\quad V_M\ =\ NO_S\uplus A_S\uplus V'\ =\ A^*_S\uplus NO^*_S\uplus V'$$
Recalling that in each graph the macls are the complements of the mvcs it holds that
$$MACL(G)=\big\{V\setminus Z:\ Z\in MVC(G)\big\}\stackrel{T.6(b)}{=\!=}\big\{(T(G)\uplus V_M)\setminus Z:\ Z\in MVC(G)\big\}$$
$$\stackrel{T.6(a)}{=\!=}\big\{T(G)\uplus (V_M\setminus Z):\ Z\in MVC(G)\big\}
\stackrel{T.7(i)}{=\!=}\big\{T(G)\uplus (V_M\setminus (A_S\uplus Z')):\ Z'\in MVC(G[V'])\big\}$$
It follows from from (30) (see also Figure 15) that the sets $V_M\setminus(A_S\uplus Z')$ are exactly the sets $A_S^*\uplus X'$ when $X'$ ranges over  $MACL(G[V'])$.  $\square$

 \begin{center}
\includegraphics[scale=0.5]{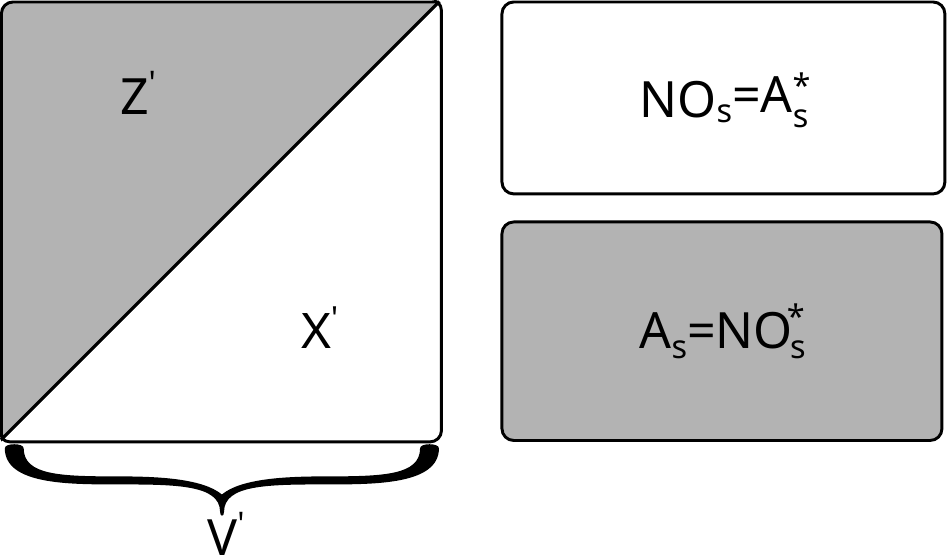}
\end{center}
{\tt Figure 15: Visualizing the proof of Theorem 8}

 Like Theorem 7 also Theorem 8 supports compressed enumeration. Specifically, how can $MACL(G[V'])$ in Theorem 8 be compressed? First off, $MACL(G[V'])=\{V'\setminus Z':\ Z'\in MVC(G[V'])$. Recall from 11.6 that (modulo trivial manipulations with 01g-rows) $MVC(G[V'])$ can be exhausted by 012-rows $r$ that derive from a certain Boolean 2-CNF. Therefore, if $r\s MVC(G[V'])$ and $r^c:=\{V'\setminus Z':\ Z'\in r\}$ then $r^c\s MACL(G[V'])$. Pleasantly $r^c$ can again be written as 012-row, For instance $r=(2,0,1,1,2,2,0)\Ra r^c=(2,1,0,0,2,2,1)$.

{\bf 11.8} The enumeration of all macls of a  bipartite (not  a general KE) graph $B$ was previously obtained in [KMNF,Thm.2] It inspired (among other factors) Theorem 8. Let us point out differences and similarities.
To begin with, also [KMNF,Thm.2] hinges on a initial fixed maximum matching $M$. In a "First reduction" $B$ gives way to some subgraph $B'$ with  perfect matching that corresponds to our subgraph $B[V']$. While we proceeded via the allmacl shadow $A^*_S$, in [KMNF] the subgraph $B'$ is pinned down by looking at alternating paths that start in the tail $T(B)$ (other terminology is used). In a "Second Reduction" the macls of $B'$ are enumerated by considering, like the author, an auxiliary digraph and its strongly connected components. Unfortunately (contrasting 9.6, 9.7) the authors  [KMNF] never proceed to the more comfortable level of posets and ideals, let alone to the {\it compressed} enumeration of all ideals.

{\bf 11.9} It follows from Theorem 8  that the core $A^*(G)$ of each KE graph $G=(V,E)$ decomposes as $A^*(G)=T(G)\uplus A_S^*\uplus A^*(G[V'])$. Yet calculating $A^*(G[V'])$, or more generally $A^*(H)$ of any KE graph $H$ with  perfect matching, is subtle and work in progress\footnote{The key is that $A^*(H)\uplus NO^*(H)$ corresponds to the   "rigid ideal" associated to every pure 2-CNF in Section 3 of [W5]. Article [W5] is an unpublished predecessor of [W4] which contains worthwile material not covered in [W4].}. Again the bipartite case  is more tractable:

{\bf Theorem 9: }{\it For each bipartite graph $B=(V,E)$ its core  $A^*(B)$ equals $T(B)\uplus A_S^*(B)$. Here $A_S^*$ is the allmacl-shadow of the tail $T(B)$.}

{\it Proof.} As noted above, $A^*(B)=T(B)\uplus A_S^*\uplus A^*(B[V'])$. The induced graph $H:=B[V']$ stays bipartite and has a perfect matching. Therefore the two shores of $H$ are {\it disjoint} macls, which implies that $A^*(H)=\es$.  $\square$

\section{From 012ac-rows to bipartite-valued rows}

In Sections 7 and 8 we investigated ways to improve the standard (a,c)-algorithm. Here we do the same but along other lines. Both approaches have their pros and cons which will be discussed at the end of this Section.
All of this not only concerns $Acl(G)$ but also  $MACL(G)$.

{\bf 12.1} 
Consider  $G_8=(V,E)$ from Figure 16a. Subjecting vertex $\alpha$ to the standard (a,c)-algorithm (say wrt the vc $Z=\{\alpha,b,\gamma,d,e\}$) yields the (a,c)-wildcard $(a_1,c_1,..,c_1)$ indexed by the elements of $star(\alpha)$ (shown in Fig.16b). Turning to $star(b)$ we find that its wildcard $(a_2,c_2,c_2,c_2)$ clashes with the first wildcard in a {\it severe} way. By definition "severe" means that the graph induced by $star(\alpha)\cup star(b)$ is no longer bipartite since it contains the triangle $\{\alpha,b,i\}$ (Fig.16c). We therefore postpone\footnote{So for the time being we  merely strive to pack the anticliques of $G_8[V\setminus\{b\}]$. Accordingly vertex $b$ will be absent in Figures 16d and 16e.} vertex $b$,  
and turn to $\gamma\in Z$. While its wildcard $(a_3,c_3,c_3,c_3)$ again clashes with $(a_1,c_1,..,c_1)$, this is not severe since the graph induced by $star(\alpha)\cup star(\gamma)$ stays bipartite (Fig.16d).
In this way piling up stars as long as possible we can  add $star(d)$ (Fig.16e) and $star(e)$ (Fig.17c). 

\begin{center}
\includegraphics[scale=0.9]{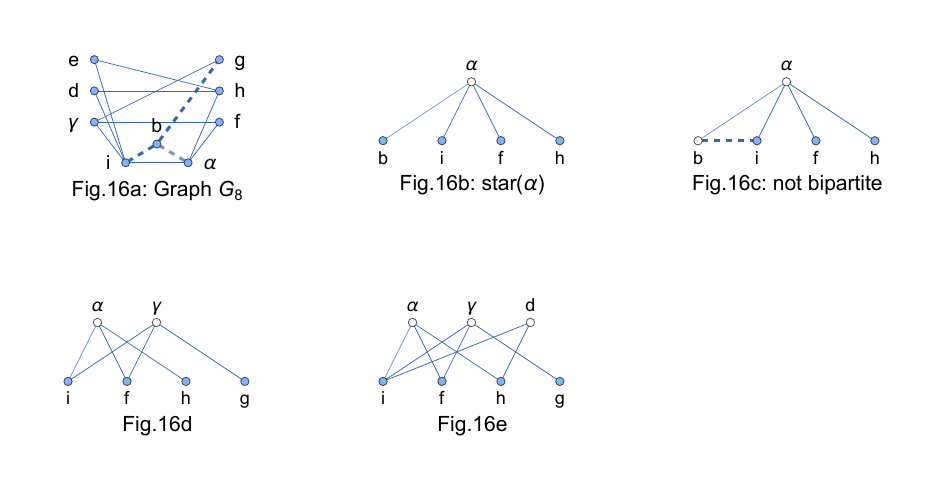}
\end{center}

What about the postponed vertex $b\in Z$? Evidently, the anticliques $Y\in Acl(G_8)$ with $b\not\in Y$ are exactly the anticliques of the bipartite graph $H(r_1'''')$  in Fig.17c (its strange name  will make sense soon). On the other hand, take any $Y\in Acl(G_8)$ with 
$b\in Y$. In view of Figure 16a it follows that $\alpha,g,i\not\in Y$. If we remove these vertices (and all incident edges) from Fig.17c, we obtain Fig.17f. Hence the sets $Y\in Acl(G_8)$ with $b\in Y$ are exactly the sets $\{b\}\uplus X$ when $X$ ranges over $Acl(H(r_2'''')$.

\begin{center}
\includegraphics[scale=1.1]{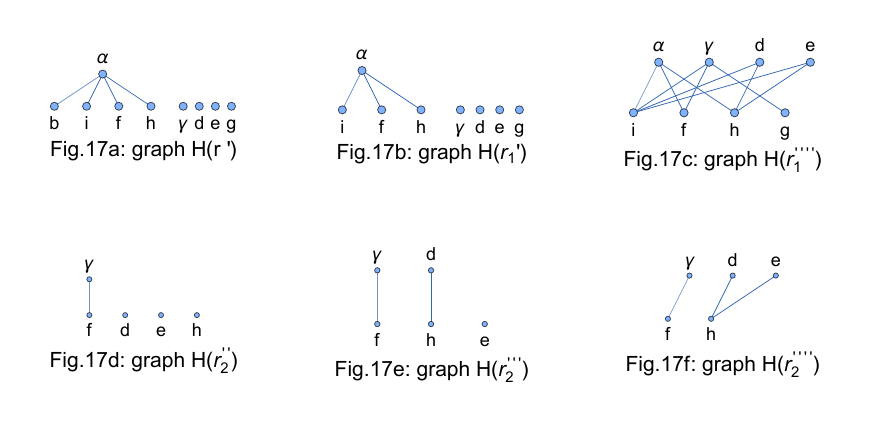}
\end{center}

{\bf 12.2} Here come the technical details  of the glimpsed relaxation of the  (a,c)-algorithm. We  define a {\it bipartite-valued row} on a set $V$ as a triplet 
$r = (zeros(r), ones(r), H(r))$
 such that $zeros(r)$ and $ones(r)$ are disjoint subsets of $V$ and such that $H(r)$ is  any bipartite graph with vertex-set $V \setminus (zeros(r) \uplus ones(r))$. By definition $r$ {\it represents} the set system 
 
 $$\quad [r]:=\Big\{ones(r) \uplus X:\ X\in Acl(H(r))\Big\}.$$ 

 While $zeros(r)\s V$ is uniquely determined by $ones(r)$ and the vertex-set of $H(r)$, the subset $zeros(r)$ is important enough to be explicitly mentioned  in the definition of a bipartite-valued (bip-val) row. As to the don't-care 2's in old-fashioned  012ac-rows, formally they have disappeared  but they survive as the isolated vertices of $H(r)$. One sees that each 012ac-row $r$ can be viewed as a bip-val row
 $(zeros(r),ones(r),H(r))$ whose bipartite graph $H(r)$ is a simple kind of bipartite graph, i.e. a galaxy.
 The new method, called {\it bip-val algorithm}, will represent $Acl(G)$ as disjoint union of 
 sets $[\rho]$, where the $\rho$'s are certain (final) bip-val rows. To achieve this we launch (akin to Section 2) a LIFO-stack of (preliminary) bip-val rows (Table 13).
 Here comes again $G:=G_8$, eager to be dealt with in more systematic fashion.

\begin{tabular}{|c|c||c} 
 \hline
 $r=$ & $\Big(\es,\ \es;\  \langle \alpha,b,..,i\rangle\Big)$ & pending $\alpha\to\ol{b,f,h,i}$\\ \hline
 & &\\ \hline
 $r'=$ & $\Big(\es,\ \es;\  H(r')\Big)$ & pending $b\to\ol{\alpha,g,i}$\\ \hline
 & &\\ \hline
  $r_1'=$ & $\Big((\{b\},\ \es;\  H(r_1')\Big)$ & pending $\gamma\to\ol{f,g,i}$\\ \hline
   $r_2'=$ & $\Big(\{\alpha,g,i\},\ \{b\};\  \langle \gamma,d,e,f,h\rangle\Big)$ &
   pending $\gamma\to\ol{f,g,i}$\\ \hline
   
   &$\vdots$ & $\vdots$\\ \hline

    $r_1''''=$ & $\Big(\{b\},\ \es;\  H(r_1'''')\Big)$ & final\\ \hline
   $r_2'=$ & $\Big(\{\alpha,g,i\},\ \{b\};\  \langle \gamma,d,e,f,h\rangle\Big)$ &
   pending $\gamma\to\ol{f,g,i}$\\ \hline

   & &\\ \hline
 $r_2'=$ & $\Big(\{\alpha,g,i\},\ \{b\};\  \langle \gamma,d,e,f,h\rangle\Big)$ &
   pending $\gamma\to\ol{f,g,i}$\\ \hline
   
   & &\\ \hline
    $r_2''=$ & $\Big(\{\alpha,g,i\},\ \{b\};\  H(r_2'')\Big)$ &
   pending $d\to\ol{h,i}$\\ \hline
    & &\\ \hline
    $r_2'''=$ & $\Big(\{\alpha,g,i\},\ \{b\};\  H(r_2''')\Big)$ &
   pending $e\to\ol{h,i}$\\ \hline
    & &\\ \hline
    $r_2''''=$ & $\Big(\{\alpha,g,i\},\ \{b\};\  H(r_2'''')\Big)$ &
   final\\ \hline

\end{tabular} 

{\sl Table 13: Snapshots of the  LIFO-stack triggered by the bip-val algorithm applied to $G_8$}

While the starter row for the standard (a,c)-algorithm applied to $G=(V,E)$ is $(2,2,...,2)$, for the bip-val algorithm it is $r=(\es,\es;V)$ (Table 13). Generally $\langle j,..,k\rangle$ denotes a graph without edges and (isolated) vertices $j,..,k$.
Upon imposing the quasi-implication $\alpha\to\ol{b,f,h,i}$ (corresponding to $star(\alpha)$) upon $r$ we obtain $r'$ whose component $H(r')$ is the graph in Fig.17a. Imposing\footnote{So different from 12.1 we do not postpone the trouble vertex $b$.} the quasi-implication $b\to\ol{\alpha,g,i}$ upon $r'$ forces us (for the sake of bipartiteness) to split $r'$ into $r_1'$ and $r_2'$. See Fig.17b for $H(r_1')$ where vertex $b$ is missing since $b\in zeros(r_1')$. As to $H(r_2')$, it consists of the isolated vertices $\gamma,d,e,f,h$ (whereas $b\in ones(r_2')$ forces $\alpha,g,i\in zeros(r_2')$). 

The vertical dots in Table 13 indicate that the top row $r_1'$ in the current LIFO stack becomes $r_1'',r_1''',r_1''''$ upon imposing
$\gamma\to\ol{f,g,i}$, then $d\to\ol{h,i}$, then $e\to\ol{h,i}$. We leave it to the reader to   spell out $r_1''$ and $r_1'''$. But $r_1''''$ is spelled out in Table 13 and $H(r_1'''')$ is rendered in Fig.17c. Since $r_1''''$ is {\it final}, i.e. all quasi-implications triggered by the vertices in $Z$ have been imposed, one removes
$r_1''''$ from the LIFO stack and stores it elsewhere.
Now $r_2'$, which has $\gamma\to\ol{f,g,i}$ pending, is the only row in the LIFO stack. Notice that $\gamma\to\ol{f,g,i}$ boils down to $\gamma\to\ol{f}$ because $g,i\in zeros(r_2')$.
Imposing $\gamma\to\ol{f}$ yields $r_2''$ (see Fig.17d).
Similarly  $r_2''',r_2''''$ are obtained (Fig.17e,f).

We see that merely two bipartite-valued rows are necessary to pack all anticliques, i.e. $Acl(G_8)=[r_1'''']\uplus [r_2'''']$. 
(The (a,c)-algorithm requires more 012ac-rows, try.)

{\bf 12.3}  We will not provide a water-tight formal argument (as we did with Thm.1 in the case of 012ac-rows) that the toy example in 12.2 generalizes to arbitrary graphs. However, fact (31) below goes a long way in that regard. 
Akin to Sec. 2.1, saying that $t\ra\ol{C}$ {\it satisfies} a set $Y$ means that either $t\not\in Y$ or $(t\in Y\ and\ Y\cap C=\es)$. 

\begin{itemize}
    \item[(31)] Let $r = (zeros, ones, H)$ be a bip-val row on $V$, and let $t \ra \ol{C}$ be a (pending) anti-implication on $V$. Let ${\cal F}$ be the family of all $Y \in [r]$ that satisfy $t \ra \ol{C}$. Then there either is one bip-val row $r^*$ with ${\cal F} =  [r^*]$, or there are two bip-val rows $r_1$ and $r_2$, such that ${\cal F} =[r_1] \uplus [r_2]$.
\end{itemize} 

{\it Proof of (31).} Of course $zeros,ones, H$ in (31) abbreviate $zeros(r),ones(r),H(r)$. Three trivial cases are treated beforehand. First,
suppose we had $t\in ones$. By the nature of the bip-val algorithm this is only possible because an quasi-implication of type $t\ra\ol{D}$ was {\it previously} imposed on some bip-val row (which then triggered $r$). Necessarily $D=C=NH(t)$, and so $t\ra\ol{C}$ holds already for all $Y\in [r]$. (In fact the algorithm does not ask in the first place that $t\ra\ol{D}$ be imposed again.). Second, suppose there was $s\in C\cap ones$. Then some quasi-implication $s\ra\ol{D}$ must have been imposed already, which in turn implies that $t\in zeros$ (because $s,t$ are adjacent). But then $t\ra\ol{C}$ holds already for all $Y\in [r]$.  Third, if either $t\in zeros$ or $C\s zeros$, then $t\ra\ol{C}$ holds already for all $Y\in [r]$, and so $r^*:=r$ does the job. 

Because of the above we can assume that $t\not\in zeros\uplus ones$ and that $C_0:=C\setminus (zeros\uplus ones)$ is nonempty. The following graph $H^*$ is therefore well-defined. It extends $H$ in that it has the additional edges $\{t,v\}\ (v\in C_0)$.

{\bf Case 1:} $H^*$ remains bipartite (like $H$). We claim that 
$$r^*:=(zeros,ones,H^*)$$
satisfies $[r^*]={\cal F}$. To fix\footnote{The reader is invited to imagine slightly more complex bip-val rows and convince himself that things keep on going smoothly.} ideas, suppose that $r$ in (31) is $r:=r_2''=\big(\{\alpha,\gamma,i\},\{b\},H(r_2'')\big)$ in Table 13 (Fig.17d), and that the pending quasi-implication
$t\ra\ol{C}$ is $d\ra\ol{h,i}$. Because $i\in zeros$, by definition $H^*$ arises from $H=H(r_2'')$ by 
adding the single edge $\{d,h\}$. Therefore $H^*=H(r_2''')$ (Fig.17e), 
and so by definition $r^*=\big(\{\alpha,\gamma,i\},\{b\},H(r_2''')\big)$. Let us check that ${\cal F}\s [r^*]$, and afterwards that $[r^*]\s {\cal F}$.

As to the claim ${\cal F}\s [r^*]$, fix any $Y\in {\cal F}$. Hence by definition $Y=\{b\}\uplus X$ for some acl of $H(r_2'')$ and $Y$ satisfies $d\ra\ol{h,i}$. {\it Subcase 1:} $d\not\in X$ (say $X=\{\gamma,e,h\}$). Then
$X$ remains an acl in $H(r_2''')$ since the new edge $\{d,h\}$ is not contained in $X$. Therefore $Y\in [r^*]$. {\it Subcase 2:} $d\in X$. Since by assumption $Y=\{b\}\uplus X$ satisfies $d\ra\ol{h,i}$, it holds that $X\cap\{h,i\}=\es$ (say $X=\{d,e,f\}$). Hence again $X$ remains an acl $H(r_2''')$, and so $Y\in [r^*]$.

As to the claim $[r^*]\s {\cal F}$, pick $Y=\{b\}\uplus X$  from $[r^*]$. So $X$ is an acl of $H(r_2''')$.
A fortiori $X$ is an acl of $H(r_2'')$ which has fewer edges. Further $X$ satisfies $d\ra\ol{h,i}$ because $i\in zeros(r_2''')$ and $\{d,h\}$ is an edge of $H(r_2''')$. (For instance $X=\{\gamma,d,e\}$.) It follows that $Y\in {\cal F}$.

{\bf Case 2:} $H^*$ is no longer bipartite. As postulated in (31) we show  that ${\cal F}=[r_1]\uplus[r_2]$ for suitable bip-val rows $r_1,r_2$. Roughly speaking (i.e. leaving aside the 0's and 1's in bip-val rows) $r_1$ will have the property that $H(r_1)$ arises from $H(r)$ by removing vertex $t$ (and thus all edges $\{t,v\}$). And $H(r_2)$ will be even smaller in that the whole closed neighborhood $NH[t]$ is removed from $H(r)$. In particular, both $Acl(H(r_1))$ and $Acl(H(r_2))$ will be contained in $Acl(H(r))$.

To fix ideas, suppose that $r$ in (31) is $r:=r'=(\es,\es,H(r'))$ in Table 13 (Fig.17a), and that the pending quasi-implication $t\ra\ol{C}$ is $b\ra\ol{\alpha,g,i}$. We then define $r_1,r_2$ as 
$$r_1:=r_1'=(\{b\},\es,H(r_1'))\ \ and\ \ r_2:=r_2'=(\{\alpha,g,i\},\{b\},H(r_2')),$$
where $H(r_2')=\langle \gamma,d,e,f,h\rangle$ (Table 13). Recall that $\cal F$ is defined as the family of all $Y\in [r']$  that satisfy $b\ra\ol{\alpha,g,i}$ (i.e. $t\ra\ol{C}$). In order to establish ${\cal F}=[r_1']\uplus[r_2']$, we first show $\s$, then $\supseteq$.

As to ${\cal F}\s[r_1']\uplus[r_2']$, pick $Y\in{\cal F}$. So in particular $Y\in[r']$, i.e. $Y=\es\uplus Y$ is some acl of $H(r')$. {\it Subcase 1:} $b\not\in Y$. Then $Y\in [r_1']$ because the acls of $H(r_1')$ are exactly the acls of $H(r')$ not containing $b$.
{\it Subcase 2:} $b\in Y$.
Because $Y$ additionally satisfies $b\ra\ol{\alpha,g,i}$ by definition of $\cal F$, it holds that $Y=\{b\}\uplus X$ for some acl $X$ of $H(r')$ satisfying $X\cap\{\alpha,g,i\}=\es$. It follows that $Y\in[r_2']$. 

As to $[r_1']\uplus[r_2']\s{\cal F}$, first pick $Y\in[r_1']$. Then $Y=\es\uplus Y$ is an acl of $H(r_1')$, hence a fortiori an acl of $H(r')$, hence $Y\in [r']$. Furthermore $Y$ satisfies 
$b\ra\ol{\alpha,g,i}$ because $b\in zeros(r_1')\Ra b\not\in Y$. Therefore $Y\in{\cal F}$. Now pick $Y\in [r_2']$. Then $Y=\{b\}\uplus X$ for some acl $X$ of $H(r_2')$, which a fortiori is an acl of $H(r')$. Hence $Y\in [r']$. Furthermore, $\{\alpha,g,i\}=zeros(r_2')$ implies that $Y\cap\{\alpha,g,i\}=\es$, hence $Y$ satisfies $b\ra\ol{\alpha,g,i}$, hence $Y\in{\cal F}$. $\square$

 {\bf 12.4} Here come  eight algorithms  that elevate the standard (a,c)-algorithm from galaxies to general bipartite subgraphs.  Whenever an odd-cycle-cover (occ) $V_1$ is involved then $V_2$ is always defined as $V_2:=V\setminus V_1$, and so $G[V_2]$ is bipartite. As previously mentioned, it seems best when $V_1$ has one of two extreme shapes. Either $V_1$ is a clique (which is achievable the better the denser $G$ is), or $V_1$ is an anticlique (which is achievable the better the sparser $G$ is). In the sequel we explore the subtler anticlique scenario.

The eight algorithms can be classified w.r.t. three criteria. The first criterion distinguishes acls (algorithms A,B,C,D below) and macls (E,F,G,H). The second  concerns counting objects (A,C,E,G) versus listing objects (B,D,F,H). The third concerns odd-cycle-covers (A,B,E,F) versus bipartite-valued rows (C,D,G,H). Put another way, the four algorithms on the left in Table 14 involve only {\it one} bipartite graph (i.e. $G[V_2]$), whereas the four algorithms on the right involve {\it several} bipartite graphs $H(r_i)$.  On the plus side, they escape all $\cal L$-related burden.

 \begin{tabular}{ | m{1.5cm} || m{3.6cm}| m{3.1cm} | | m{3.3cm}| m{3cm} |  } 
  \hline
   Ordinary anticliques:
   & {\bf A. Counting all acls} using an indpt occ $V_1$: Compress $Acl(G[V_2])$ with $t'$ many 012ac-rows and combine this with 
   $\cal L$-related information to get $ms(G)=|Acl(G)|$.
   
  & {\bf B. Listing all acls} using an  indpt. occ: Similar to {\bf A} but {\it more} $\cal L$-related information is necessary.
  
  & {\bf C. Counting all acls} using bip-val rows $r_i\ (i\le R)$: Apply the algorithm of [CT] to all bip. graphs $H(r_i)$ and add up the results. 
  
  & {\bf D. Listing all acls} with bip-val rows $r_i$: Similar to {\bf C} but handling the bipartite graphs $H(r_i)$ is more complex.
  \\ 
  \hline
  Maximum anticliques: 
  & {\bf E. Counting all macls} with an indpt. occ $V_1$: Similar to {\bf F}, but easier. 
  
  & {\bf F. Listing all macls} with an indpt. occ $V_1$:
  Here only $\cal L$ must be known. One gets $MACL(G)$ by a subtle interplay between $\cal L$ and the compression of $Acl(G[V_2])$. 
  
  & {\bf  G. Counting all macls} with bip-val rows $r_i$: Determine the cardinalities $|MACL(H(r_i))|$ by applying sat-count to 2-CNF's. Summing up yields $|MACL(G)|$. 
  
  & {\bf H. Listing all macls} with bip-val rows: According to 11.7 all families $MACL(H(r_i))$ can be compressed
  efficiently via 01g-rows. Their union is $MACL(G)$. 
  \\ 
  \hline 
\end{tabular}

{\sl Table 14: Bird's eye view of eight algorithms}

Additional comments are in order for the algorithms A,B,F: 
 
\begin{itemize}
 \item[\bf A.] Since $G[V_2]$ is bipartite,  the compression of $Acl(G[V_2])$ works particularly well. Thus the smaller shore $Z$ of $G[V_2]$ is a vertex-cover on which the standard (a,c)-algorithm can be based; or one can drive that further  as shown in 7.3. Whatever way, suppose one has computed 012ac-rows $r_i$ such that  $Acl(G[V_2])=r_1\uplus\cdots\uplus r_{t'}$. 
As to $V_1$, suppose ${\cal L}=\{Y_1,...,Y_t\}$ and all {\it cardinalities} $\beta_i:=|EC(Y_i)|\ (Y_i\in {\cal L})$ have been calculated\footnote{Possibly calculating the mere cardinalities can be achieved faster than calculating the equivalence classes $EC(Y_i)\ (Y_i\in {\cal L})$ themselves. Also recall the brute-force way in 8.5.}. For each $r_j$ let $r_j[i]$ be the 012ac-row obtained from $r_j$ upon filling in 0's on the positions indexed by the elements of $NH(Y_i)$, and put $\gamma(j,i):=|r_j[i]|$ (except that $\gamma(j,i):=0$ if $NH(Y_i)\cap ones(r_j)\neq\es$). Then 
$$(32)\ ms(G)=\beta_1\big(\gamma(1,1)+\cdots +\gamma(t',1)\big)+
\beta_2\big( \gamma(1,2)+\cdots +\gamma(t',2)\big)+\cdots+
\beta_t\big( \gamma(1,t)+\cdots +\gamma(t',t)  \big)$$

 \item[\bf B.] Finding $Acl(G)$ itself (not just $ms(G)$) is obviously harder than {\bf A}. Now the equivalence classes $EC(Y_i)\ (Y_i\in {\cal L})$ themselves,  preferably in a compressed format by means of disjoint 012-rows (as glimsed in 8.3, 8.4) must be produced. Each product of numbers, such as $\beta_3\gamma(7,3)$ in (32), gets replaced by a certain bundle of anticliques. Namely, suppose that
 $EC(Y_3)=\sigma_1\uplus\cdots\uplus\sigma_{68}$ for some 012-rows $\sigma_k$. If $\oplus$ is used for the concatenation of rows then the bundle of anticliques that replaces $\beta_3\gamma(7,3)$ is
 $$(\sigma_1\oplus r_7[3])\uplus (\sigma_2\oplus r_7[3])\uplus\cdots\uplus (\sigma_{68}\oplus r_7[3])$$

\item[\bf F.] Finding $MACL(G)$ is  similar to {\bf B}, but surprisingly {\it  easier} in that only $\cal L$ is needed (without the attached equivalence classes). Namely,
because each $Y_i\in{\cal L}$ is the largest member in its equivalence class $EC(Y_i)$, each macl $S$ of $G$ must  be of type $S=Y_i\uplus X$ for some $j$ and  some $X\in r_j[i]$. Let us pile up bundles of macls until $MACL(G)$ is exhausted.

{\bf Subcase 1:} $\alpha(G)$ is known.  For any 012ac-row $\rho$ recall from Sec.5 the meaning of $\alpha(\rho)$
and $MACL(\rho)$. Evidently for each $1\le i\le t$ and each $1\le j\le t'$ it holds that
$$|Y_i|+\alpha(r_j[i])\le\alpha(G),$$
and that calculating $\alpha(r_j[i])$ is lightning fast.  Whenever $\le $ is $=$, add the macl-bundle $Y_i\oplus MACL(r_j[i])$.
Because "=" may take place very rarely, this algorithm is not polynomial in the size of $MACL(G)$.

{\bf Subcase 2:} $\alpha(G)$ is not known. Then  keep a record of the maximum value of $|Y_i|+\alpha(r_j[i])$ achieved thus far and of the pairs $(i,j)$ that achieve this value. After scanning all pairs $(i,j)$  one knows both $\alpha(G)$ and the pairs $(i,j)$ that yield macl-bundles $Y_i\oplus MACL(r_j[i])$. 

Unfortunately even when $\alpha(G)$ is {\it known beforehand}, one still needs to scan all pairs. Hence the effort in Subcase 1 and 2 is about the same! Could it be that in  situations where $t\cdot t'$ is small, the mere scanning of $t\cdot t'$ pairs $(i,j)$ is a competitive algorithm for calculating $\alpha(G)$ in the first place?  (Algorithm {\bf G} can also be adapted to just calculate $\alpha(G)$. That is most tempting for forest-valued rows, see 12.7.)

\end{itemize}

 {\bf 12.5}  The  coding of the bip-val algorithm (work in progress) is actually {\it easier} than the coding of the (a,c)-algorithm (which can be gleaned from the proof of Theorem 1). For instance, the intricacies of Case 3 in Theorem 1 arise because the quasi-implication $t\ra\ol{C}$ to be imposed on $r$ touches upon several connected components of the galaxy $H(r)$. That can be avoided in the bip-val algorithm, {\it provided} the bipartite graphs $H(r_i)$ occuring in the bip-val rows $r_i$ are encoded in the Wolfram Language (or some other high-level programming language).
  Then not only the edges wrapped in $t\ra\ol{C}$ can be added at once to $H(r_i)$ and yield a new graph $H$, but the single Mathematica command $\tt BipartiteGraphQ[H]$ settles at once whether $\tt H$ remains bipartite.

{\bf 12.6} There is some middle ground between 012ac-rows and bip-val rows. Namely,  {\it forest-valued} rows are bip-val rows $r$ whose bipartite graphs $H(r)$ are forests. Since galaxies are forests, forest-valued rows also generalize 012ac-rows. Let us compare the "bipartite-level" with the "forest-level"; first w.r.t.  acls, then w.r.t. macls.

Concerning acls, in 7.3 and 7.4 we saw how the parameter $t'$ in the $(V_1,V_2)$-approach can be trimmed when $G[V_2]$ is bipartite. That gets better still when $H=G[V_2]$ is a forest because then $Aux(H)$ in 7.4 gives way to more efficient methods (work in progress). This carries over to the right part of Table 14, i.e. $Acl(H(r_i)$ is better compressible for forest-valued rows $r_i$ than for bip-val rows $r_i$.

Concerning macls, recall from 11.7 that for any bipartite graph $B$ with a perfect matching one can compress $MACL(B)$ with 012-rows. Here $B$ being a forest does not (in obvious ways) accelerate the calculation. For arbitrary bipartite graphs $H$ Theorem 8 reduces the calculation of $MACL(H)$ to the perfect matching case. However, this reduction relies on finding a maximum matching $M$ of $H$. Is it faster to find $M$ when $H$ is a forest? Yes indeed: While finding $M$ for a bipartite $H=(V,E)$ costs $O(\sqrt{|V|} E)$, for forests $H$ it merely amounts to peel the dangling edges of $H$ until one gets the empty tree. As above the advantage of trees concerns both the left and right half of Table 14. 

{\bf 12.7} In a forthcoming article the author plans to implement, in the Wolfram Language, some of the algorithms in Table 14, both on the forest-level and the bipartite-level. Notice that the bip-val algorithm readily adapts to forest-level: The Mathematica command $\tt BipartiteGraphQ[H]$ in 12.5 gives way to {\tt FindCycle[H]}, which decides when $H(r)$ has ceased to be a forest and hence $r$ must be split. Therefore the algorithms {\bf C,D,G,H} on the right in Table
14 are relatively easy to implement on both levels.

 As to the algorithms {\bf A,B,E,F} on the left in Table 14,
 recall that in 8.6 we formulated a 0,1-program whose solution $V_1$ is an odd-cycle-cover of $G$, and so $G[V_2]:=G[V\setminus V_1]$ is bipartite. If one prefers a forest $G[V_2]$ then this requires a very different 0,1-program. On top of that recall that in both scenarios  preferably $V_1$ should be either a clique  or an anticlique.


\section{References}

\begin{enumerate}

\item[{[CH]}] Y.Crama, P.L. Hammer, Boolean functions, Enc of Math and its Appl. 142, Cambridge 2011.
\item[{[CT]}] C.C. Chen and S.K. Tan, On the order of certain finite distributive lattices, Nanta Mathematica IX (1977), 187-194
\item[{[Fa]}] J. Farley, THE “k= 1” CASE OF A PROBLEM OF
GREENE AND KLEITMAN FROM 1976: JOIN-IRREDUCIBLE ELEMENTS IN THE LATTICE OF SPERNER 1-FAMILIES, Mathematica Pannonica New Series 27 /NS 1/ (2021) 2, 95-104.
\item[{[F]}] R. Freese, An application of Dilworth's lattice of maximal antichains, Disc. Math. 7 (1974) 107-109.
\item[{[GK]}] C. Greene, D.J. Kleitman, The Structure of Sperner k-Families,
JOURNAL OF COMBINATORIAL THEORY (A) 20, 41-68 (1979).

\item[{[GV]}] A. Gainer-Dewar and P. Vera-Licona, The minimal hitting set generation problem: algorithms and computation, SIAM J. Discrete Math. 31 (2017) 63–100.

\item[{[H]}] F. H\"uffner, Algorithm engineering for optimal graph bipartization, Journal of graph algorithms and applications (2009) 77-98. 
\item[{[KO]}] SO Kuznetsov, S. Obiedkov,
Comparing performance of algorithms for generating concept lattices, Journal of Experimental and Theoretical Artificial Intelligence Vol. 14, 189-216, 2002. 
\item[{[KMNF]}] T Kashiwabara, S Masuda, K Nakajima, T Fujisawa,  Generation of Maximum Independent Sets
of a Bipartite Graph and Maximum Cliques
of a Circular-Arc Graph, JOURNAL OF ALGORITHMS 13, 161-174 (1992).
\item[{[LM]}] V.E. Levit, E. Mandrescu, The independence polynomial of a graph-a survey, Proceedings of the 1st Internat. Conf. on Algebraic Informatics, 233-254, Thessaloniki 2005.
\item[{[M]}] V. Minasyan, On the structure of maximum independent sets in bipartite graphs, Information Theories and Applications 17 (2010) 176-187.


\item[{[S]}] A. Schrijver, Combinatorial Optimization, Springer 2003. 

\item[{[WG]}] S. Wagner, I. Gutman, Maxima and minima of the Hosoya index and the Merrifield-Simmons-index, Acta Appl. Math. 112 (2010) 323-346.
\item[{[W1]}] M. Wild, Output-polynomial enumeration of all fixed-cardinality ideals of a poset, respectively all fixed-cardinality subtrees of a tree, Order (2014) 31:121-135.
\item[{[W2]}] M. Wild, Compression with wildcards: All exact or all minimal hitting-sets, Open Mathematics 2023; 21: 20220596.
\item[{[W3]}] M. Wild, Compression with wildcards: Abstract simplicial complexes, Quaestiones Mathematicae 46 (2023) 1151-1173.
\item[{[W4]}] M. Wild, Compression with Wildcards: All Models of a Boolean 2-CNF. Contemporary Mathematics 6(6):7795-816 (2025).

\item[{[W5]}]  M.Wild, Revisiting the enumeration of all models of a Boolean 2-CNF, arXiv:1208.2559v5 (2016).
\item[{[WYL]}]  K. Wang, K. Yu, C. Long, Efficient k-Clique Listing: An Edge-Oriented Branching Strategy,
Proceedings of the ACM on Management of Data, Volume 2, Issue 1 (2024).
\item[{[YLDCW]} ] X. Ye, R.H. Li, Q. Dai, H. Chen, G. Wang, Lightning Fast and Space Efficient k-clique Counting, Proceedings
of the ACM, Web Conference April 25–29, 2022, Virtual Event.
\item[{[Z]}] J.Zito, The structure and maximum number of maximum independent sets in trees, Journal of Graph Theory 15 (1991) 207-221.
\end{enumerate}

\end{document}